\documentclass[11pt]{article}

\usepackage{lineno}

\modulolinenumbers[1]
\usepackage{pstricks}

\newcommand{\lb}{\left(}
\newcommand{\rb}{\right)}

\usepackage{amsmath}
\usepackage{amssymb}
\usepackage{color}
\usepackage{soul}
\usepackage{graphicx}

\usepackage{setspace}

\begin{document} 

%%\iffalse %%

\begin{center}
{\LARGE {\bf A Non-Hydrostatic Multi-Phase Mass Flow Model}}
\\[6mm]
{\large {\em Shiva P. Pudasaini}\\[2mm]
Technical University of Munich, Chair of Landslide Research}\\
{Arcisstrasse 21, D-80333, Munich, Germany}\\[1mm]
{E-mail: shiva.pudasaini@tum.de}\\[7mm]
\end{center}
{\bf Abstract:} Modeling mass flows is classically based on the hydrostatic, depth-averaged balance equations. However, if the momentum transfers scale similarly in the slope parallel and the flow depth directions, then the gravity and the acceleration can have the same order of magnitude effects. This urges for a non-hydrostatic model formulation. Here, we extend existing single-phase Boussinesq-type gravity wave models by developing a new non-hydrostatic model for multi-phase mass flows consisting of the solid and fine-solid particles, and viscous fluid (Pudasaini and Mergili, 2019). The new model includes enhanced gravity and dispersion effects taking into account the interfacial momentum transfers due to the multi-phase nature of the mass flow. We outline the fundamentally new contributions in the non-hydrostatic Boussinesq-type multi-phase gravity waves emerging from the phase-interactions including buoyancy, drag, virtual mass and Newtonian as well as non-Newtonian viscous effects. So, this contribution presents a more general, well-structured  framework of the multi-phase flows with enhanced gravity and dispersion effects, setting a foundation for a comprehensive simulation of such flows. We discuss some particular situations where the non-hydrostatic and dispersive effects are more pronounced for multi-phase mass flows. Even the reduced models demonstrate the importance of non-hydrostatic contributions for both the solid and fine solid particles, and the viscous fluid. Analytical solutions are presented for some simple situations demonstrating how the new dispersive model can be reduced to non-dispersive motions, yet largely generalizing the existing non-dispersive models. We postulate a novel, spatially varying dissipative force, called the prime-force, which physically controls the dynamics, run-out and the deposition of the mass flow in a precise way. The practitioners and engineers may find this force very useful in relevant technical applications. This illuminates the need of formally including the prime-force in the momentum balance equation. A simple dispersion equation is derived. We highlight the essence of dispersion on the mass flow dynamics. Dispersion consistently produces a wavy velocity field about the reference state without dispersion. Emergence of such a dispersive wave is the first of this kind for the avalanching debris mass. We reveal that the dispersion intensity increases energetically as the solid volume fraction or the friction decreases.

\section{Introduction}

Natural hazards associated with geophysical mass movements consist of a mixture of granular materials of different sizes of particles and the fluid with their respective physical properties. There have been rapid advancements in modeling shallow granular material (Savage and Hutter, 1989; Denlinger and Iverson, 2001; Pouliquen and Forterre, 2002, McDougall and Hungr, 2005; Pudasaini and Hutter, 2007; Luca et al., 2009; Kuo et al., 2011) and particle fluid mixture (Iverson, 1997; Pitman and Le, 2005; Pudasaini, 2012; Pudasaini and Mergili, 2019) mass flow modelling. These models are primarily based on the hydrostatic pressure assumptions. However, due to the centrifugal acceleration, the mass flows in curved channels also include some non-hydrostatic contributions to hydraulic pressure gradients and the Coulomb friction forces because of the enhanced normal load (Savage and Hutter, 1991; Gray et al., 1999; Pudasaini et al., 2005). Furthermore, Pailha and Pouliquen (2009), and Pudasaini (2012) showed that the pressure in mixture mass flows can be non-hydrostatic due to the Newtonian and non-Newtonian viscous contributions, the particle concentration distributions, and the relative velocity between particle and fluid.  
\\[3mm]
Classically, modeling geophysical flows is usually based on the hydrostatic, depth-averaged mass and momentum balance  equations. Hydrostatic flow models are based on the assumption that the slope parallel length scale is much larger than the length scale in the flow depth direction. However, if the similar length scalings are required in the slope parallel and the flow depth directions, then the gravity and the vertical acceleration can have the same order of magnitude effects (Denlinger and Iverson, 2004; Pudasaini and Hutter, 2007, Castro-Orgaz et al., 2015). This may call for the use of the full (without reducing to the hydrostatic condition) momentum equation also in the slope normal direction as in the slope parallel directions (Domnik et al., 2013). Denlinger and Iverson (2004) first mentioned that the vertical accelerations in granular mass flows can be of the same order of magnitude as the gravity acceleration. In this situation, the vertical acceleration can be as significant as the acceleration in the slope parallel direction. This is particularly so for steep, irregular and curved slopes where there is a substantial acceleration of the material in the flow depth direction. So, they suggested for the non-hydrostatic modeling of granular flows. This requires enhancements of the normal stress (in the slope normal or vertical direction) that results in the enhancements of the acceleration, friction and fluxes in the momentum balance equations (Castro-Orgaz et al., 2015; Yuan et al., 2018). Since the Coriolis acceleration is usually neglected in geophysical mass flows (such as landslide and avalanches), two-types of accelerations can be induced affecting the normal stress distributions of the free surface flows (Castro-Orgaz et al., 2015): First, the acceleration due to the real forces acting at the bed-normal direction. Second, the centripetal acceleration that arises due to the curved flow path (Savage and Hutter, 1991; Pudasaini and Hutter, 2007). The first is the main contributor of the Boussinesq-type models, while both combined result in the more comprehensive non-hydrostatic flows. 
\\[3mm]
Following the work of Boussinesq (1872, 1877), the free surface water flow simulations are generally based on non-hydrostatic depth-averaged models. Fundamental further contributions in including Boussinesq-type non-hydrostatic and dispersive effects in water waves are also due to Serre (1953), Peregrine (1967) and Green et al. (1976). The recent advancements, analyses and applications of the dispersive wave characteristics of the Boussinesq system with advanced numerical schemes for real flow simulations include the works by 
 Nwogu (1993), Wei and Kirby (1995), Madsen and Sch\"affer (1998), Kennedy et al. (2000), 
Stansby (2003), Chen et al. (2003), Erduran et al. (2005), Chen (2006), and Kim and Lynett (2011). For detailed review on it, we refer to Castro-Orgaz et al. (2015). However, for shallow granular flows the effect of nonzero vertical acceleration on depth-averaged  momentum fluxes and stress states were first included by Denlinger and Iverson (2004) while modeling granular flows across irregular terrains. This was later extended by Castro-Orgaz et al. (2015) resulting in the novel Boussinesq-type theory for granular flows. Utilizing the non-hydrostatic Boussinesq-type water wave theory, Castro-Orgaz et al. (2015) rigorously developed a non-hydrostatic depth-averaged granular flow model. Considering the vertical motion of particles, they explicitly determined the vertical velocity, vertical acceleration, and vertical normal stresses from the mass and momentum conservation equations. They have shown that granular mass flow can be described by fully non-linear, Boussinesq-type gravity waves, generalizing the basic Boussinesq-type water wave theory used in civil and coastal engineering to granular mass flows. Later, Yuan et al. (2018) advanced further by presenting a refined and more complete non-hydrostatic shallow granular flow model. They also cast their model in to a usual Boussinesq-type water wave equations. 
\\[3mm]
In developing the non-hydrostatic Boussinesq-type gravity wave models for granular flows, both Castro-Orgaz et al. (2015) and Yuan et al. (2018) considered the vertical momentum equation, assuming the shallowness of the flow depth and the constant velocity profiles of the horizontal velocity components. Along with these assumptions, there are three key aspects in their model development: Obtaining the vertical normal stress component from the vertical momentum equation, an expression for the vertical velocity component in terms of the horizontal mass flux (divergence), and the definition of the depth integration of the vertical velocity component from a generic elevation to the free surface. Finally, the depth averaged mass and momentum equations, together with these three considerations lead to a non-hydrostatic Boussinesq-type gravity wave models for granular flows. However, all these formulations are primarily based on the global horizontal-vertical Cartesian coordinate for a single-phase granular flows. 
\\[3mm]
One- and two-phase models cannot appropriately represent many important aspects of very complex mass flows in terms of material composition and interactions among the involved phases. The rheological properties and flow dynamics are governed by coarse and fine solids, and viscous fluid, i.e., typically three phases (Iverson, 1997; Pierson, 1970, 2005; de Haas et al., 2015, 2016; Pudasaini and Fischer, 2020b). Consequently, the most complex model family for geophysical mass flows should aim at describing the flow as (typically) a three-phase mixture, as often observed in the field and experiments (Major and Iverson, 1999; McArdell et al., 2007; Conway et al., 2010; Schneider et al., 2011; Johnson et al., 2012; de Haas et al., 2015, 2016; Steinkogler et al., 2015). {In general terms, the mechanical components} in the mixture mass flow can be divided into three constituents: The fluid phase is a mixture of water and very fine particles (clay, silt, colloids), the fine-solid phase consists of sand and particles larger than clay and silt, and the solid phase represents the coarse material. These materials can be described as viscoplastic, Coulomb-viscoplastic, and Mohr-Coulomb continuum.  
With this, Pudasaini and Mergili (2019) proposed a novel multi-phase, multi-mechanical mass flow model, by extending the two-phase viscous fluid and Coulomb solid model (Pudasaini, 2012) to additionally combine it with the fine-solid material. The Pudasaini and Mergili (2019) model can accurately simulate complex cascading multi-phase natural events (Mergili and Pudasaini, 2021; Mergili et al., 2020a, 2020b; Shugar et al., 2021). 
\\[3mm]
Here, we extend and utilize the above mentioned ideas to the multi-phase mass flow model (Pudasaini and Mergili, 2019) to generate a new non-hydrostatic Boussinesq-type gravity wave model for multi-phase mass flows in a locally inclined Cartesian coordinate system (Pitman and Le, 2005; Pudasaini, 2012). The new non-hydrostatic multi-phase mass flow model includes enhanced gravity and dispersion effects as in the single-phase models by Denlinger and Iverson (2004), Castro-Orgaz et al. (2015) and Yuan et al. (2018). But, our new model further includes interfacial momentum transfers in the non-hydrostatic Boussinesq-type model formulation representing the complex multi-phase nature of the mass flow. 
We delineate the fundamentally new contributions in the Boussinesq-type gravity waves in mass flows emerging from the phase-interactions. This includes buoyancy, drag, virtual mass and Newtonian plus non-Newtonian viscous effects.
We outline the first-ever application potential of the dispersive multi-phase mass flows. As in the effective gravity, the dispersive terms are strongly coupled, e.g., due to the interfacial drag and virtual mass contributions. There are direct and strong couplings between the solid, fine-solid and the fluid components among these dispersion relations. Interfacial drags bring completely new mechanisms in the non-hydrostatic, dispersion relations. 
 We discuss some particular situations where the non-hydrostatic dispersive effects are more pronounced in multi-phase particle-fluid mixture mass flows than in single-phase flows. 
 So, this contribution sets a foundation for a more comprehensive and general frame for the simulation of dispersive, multi-phase mass flows.
We also present simplified models that might be helpful in solving the equations with reduced complexity.
The reduced models already appeared to be the important generalizations and extensions of some mass flow models available in the literature. 
We formally postulate a new, spatially varying dissipative force, called the prime-force, which can physically precisely control the mass flow dynamics, run-out and the deposition. We present a simple dispersion model and its solution. Dispersion produces a wavy velocity field about the reference state without dispersion. The dispersion increases greatly as the solid volume fraction or the basal friction decreases. These are new understandings for the motion of a dispersive landslide. 

\section{Construction of the Model}

\subsection{Non-Hydrostatic Contributions}

Following Pudasaini and Mergili (2019) and Pudasaini and Fischer (2020a), first, we define the variables and parameters. 
Let the solid, fine-solid and fluid phases be denoted by the suffices $s$, $fs$, $f$, respectively. The fluid phase is {governed} by its true density $\rho_f$, viscosity $\eta_f$, and isotropic stress distribution; the fine-solid and solid phases are characterized by their true densities $\rho_{fs}$, $\rho_s$; internal friction angles $\phi_{fs}$, $\phi_{s}$; basal friction angles $\delta_{fs}$, $\delta_{s}$; and anisotropic stress distribution, $K_{s}$ (lateral earth pressure coefficient); and the viscosity of the fine-solid $\eta_{fs}$. Furthermore,
$\gamma_s^f = \rho_f/\rho_s$,
$\gamma_s^{fs} = \rho_{fs}/\rho_s$, 
$\gamma_{fs}^f = \rho_{f}/\rho_{fs}$ 
are the fluid to solid, fine-solid to solid and fluid to fine-solid density ratios, $\nu_f^e$ and $\nu_{fs}^e$ are the effective kinematic viscosities for the fluid and fine-solid, $\mu_s = \tan\delta_s$ and $\mu_{fs} = \tan\delta_{fs}$ are the friction coefficients for the solid and fine solid.
Let ${\bf u}_s = \lb u_s, v_s, w_s\rb$, ${\bf u}_{fs} = \lb u_{fs} , v_{fs} , w_{fs} \rb$, ${\bf u}_f = \lb u_f, v_f, w_f \rb$, and $\alpha_s$, $\alpha_{fs}$, $\alpha_f$ denote the velocities with their components along the flow directions ($x, y, z$), and the volume fractions for the solid, fine-solid, and fluid constituents. {Similarly,}  
 $p_{fs}$ and $p_{f}$ are the pressures, ${C}_{DG}$ and ${C}_{vm}$ constitute the interfacial force densities, namely, the drags and the virtual mass forces, and $C_{DV}$ are the viscous drag coefficients. The superscript-pair represents the considered phases, e.g., ${C}_{DG}^{s,f}$ means the drag force {exerted by fluid on solid}, ${\mathcal C}$ are the virtual mass coefficients, $T_{**}$ are the components of the Cauchy stress tensor, $\jmath = 1 \, \mbox{or}\, 2$ correspond to linear or quadratic drag coefficients, $g^x, g^y, g^z$ are the components of gravitational acceleration, basal- and the free-surface of the flow are denoted by $b = b(t,x,y)$ and $s = s(t,x,y)$, and $h = s - b$ is the flow depth.

\subsubsection{Derivation of Normal Stress Components} 

The non-hydrostatic modelling framework includes two important and essential components: ($i$) enhanced gravity, and ($ii$) dispersive contributions (see, e.g., Castro-Orgaz et al., 2015; Yuan et al., 2018). Both emerge from the consideration of the momentum equation in the flow depth direction such that the normal component of the velocity is retained, that was neglected in simple hydrostatic model developments as discussed at Section 1. These contributions, however, are modelled in terms of the slope parallel velocity gradients or fluxes. For this, following Pudasaini (2012), and Pudasaini and Mergili (2019), first we consider the solid momentum balance in the flow depth direction:
\begin{linenomath*}
 \begin{eqnarray}
\begin{array}{lll}
\displaystyle{
\frac{\partial}{\partial t} \lb w_s 
- \gamma^f_s \mathcal C^{s,f} \lb w_f - w_s \rb
- \gamma^{fs}_s \mathcal C^{s,fs} \lb w_{fs} -  w_s \rb
\rb}\\[3mm]
\displaystyle{+\frac{\partial}{\partial x} \lb u_sw_s 
- \gamma^f_s \mathcal C^{s,f} \lb u_fw_f - u_sw_s \rb
- \gamma^{fs}_s \mathcal C^{s,fs} \lb u_{fs}w_{fs} -  u_{s}w_s \rb
\rb
}
\\[3mm]
\displaystyle{+\frac{\partial}{\partial y} \lb v_sw_s 
- \gamma^f_s \mathcal C^{s,f} \lb v_fw_f - v_sw_s \rb
- \gamma^{fs}_s \mathcal C^{s,fs} \lb v_{fs}w_{fs} -  v_sw_s \rb
\rb
}
\\[3mm]
\displaystyle{+\frac{\partial}{\partial z} \lb w_s^2 
- \gamma^f_s \mathcal C^{s,f} \lb w_f^2 - w_s^2 \rb
- \gamma^{fs}_s \mathcal C^{s,fs} \lb w_{fs}^2 -  w_s^2 \rb
\rb
}\\[3mm]
= -\lb 1 - \gamma^f_s\rb g^z-\lb
\mu_s\frac{\partial T_{zx_s}}{\partial x}
+\mu_s\frac{\partial T_{zy_s}}{\partial y}
+\frac{\partial T_{zz_s}}{\partial z}
\rb\\[3mm]
+\displaystyle{
\frac{1}{\alpha_s}\left [ 
 C_{DG}^{s,f} \lb {w}_f -{w}_s \rb|{\bf u}_f -{\bf u}_s|^{j-1}
+C_{DG}^{s,fs} \lb {w}_{fs} -{w}_s \rb|{\bf u}_{fs} -{\bf u}_s|^{j-1}
- C_{DV}^s w_s|{\bf u}_s|\alpha_s\right ]
},
\label{w_solid_momentum}
\end{array}    
\end{eqnarray}
\end{linenomath*}
where, for simplicity, $\alpha_s$ has been taken out. Note that since both $C_{DG}^{s,f}$ and $C_{DG}^{s,fs}$ contain $\alpha_s$ in their numerators (see, Appendix), appearance of $1/\alpha_s$ in (\ref{w_solid_momentum}) makes no problem. It is important to note that (\ref{w_solid_momentum}) contains the normal stress $T_{zz_s}$ from which we can construct the full description of the normal stress in the flow depth direction that includes all the essential components emerging from the flow dynamics and interfacial momentum transfers in excess to the usual hydrostatic normal load that is simply associated with the gravity load in the flow depth direction.
\\[3mm]
We define a new variable $\eta = z -b$, the relative flow depth. Then, following the procedure in Castro-Orgaz et al. (2015) and Yuan et al. (2018), integrating (\ref{w_solid_momentum}) from the generic elevation $z$ to the free surface $s$, neglecting the shear stresses, and using the tractionless condition at the free surface (Pudasaini, 2012; Pudasaini and Mergili, 2019), we obtain an expression for the normal stress in terms of $\eta$:
{\small
\begin{linenomath*}
 \begin{eqnarray}
\begin{array}{lll}
\displaystyle{
\tau_{zz_s}(\eta) = \lb 1 - \gamma^f_s\rb g^z(h-\eta)
+ \frac{\partial}{\partial t}\left [ I_s 
- \gamma^f_s \mathcal C^{s,f} \lb I_f - I_s\rb
- \gamma^{fs}_s \mathcal C^{s,fs} \lb I_{fs} - I_s\rb
\right ]} \\[3mm]
\hspace{1.2cm}
\displaystyle{
+ \nabla\cdot\left [ I_s{\bf u}_s 
- \gamma^f_s \mathcal C^{s,f} \lb I_f {\bf u}_f - I_s {\bf u}_s\rb
- \gamma^{fs}_s \mathcal C^{s,fs} \lb I_{fs} {\bf u}_{fs} - I_s {\bf u}_s\rb
\right ]}\\[3mm] 
\hspace{1.2cm}
\displaystyle{
- \left [ w_s^2 
- \gamma^f_s \mathcal C^{s,f} \lb w_f^2 - w_s^2\rb
- \gamma^{fs}_s \mathcal C^{s,fs} \lb w_{fs}^2 - w_s^2\rb
\right ]}\\[3mm] 
\hspace{1.2cm}
\displaystyle{
-\frac{1}{\alpha_s}\left [ 
 C_{DG}^{s,f}|{\bf u}_f -{\bf u}_s|^{j-1}\left [ I_f - I_s\right ]
+ C_{DG}^{s,fs}|{\bf u}_{fs} -{\bf u}_s|^{j-1}\left [ I_{fs} - I_s\right ]
- C_{DV}^s w_s|{\bf u}_s|\alpha_s I_s
\right ],}
\label{tau_zz_a}
\end{array}    
\end{eqnarray}
\end{linenomath*}
}
\hspace{-2.5mm}
where 
\begin{linenomath*}
 \begin{eqnarray}
\begin{array}{lll}
I_s = \int_z^s w_s dz',\,\,\,\,\,
I_{fs} = \int_z^s w_{fs} dz',\,\,\,\,\,
I_f = \int_z^s w_f dz';\\[3mm]
w_s = w_{b_s} - \lb \nabla\cdot {\bf u}_s\rb\eta,\,\,\,\,\,
w_{fs} = w_{b_{fs}} - \lb \nabla\cdot {\bf u}_{fs}\rb\eta,\,\,\,\,\,
w_f = w_{b_f} - \lb \nabla\cdot {\bf u}_f\rb\eta;\\[3mm]
\nabla\cdot {\bf u}_{s} = \partial u_s/\partial x + \partial v_s/\partial y, \,\,\,\,\,
\nabla\cdot {\bf u}_{fs} = \partial u_{fs}/\partial x + \partial v_{fs}/\partial y,\,\,\,\,\,
\nabla\cdot {\bf u}_{f} = \partial u_{f}/\partial x + \partial v_{f}/\partial y.
\label{tau_zz_aa}
\end{array}    
\end{eqnarray}
\end{linenomath*}
We depth-integrate $w_s$, and define $\hat I_s$ (similar structures hold for fine-solid and fluid):
\begin{linenomath*}
 \begin{eqnarray}
\begin{array}{lll}
\displaystyle{\bar w_s: = \frac{1}{h}\int_b^s w_s dz' = w_{b_s} -\lb \nabla\cdot {\bf u}_s\rb\frac{h}{2},\,\,\,\,\,
w_{b_s} = u_s\frac{\partial b}{\partial x} + v_s\frac{\partial b}{\partial y}};
\\[5mm]
\hat I_s: = \int_b^z w_s dz'= \int_b^s w_s dz' - \int_z^s w_s dz' = h \bar w_s -I_s,
\label{tau_zz_aaa}
\end{array}    
\end{eqnarray}
\end{linenomath*}
where $b$ is the basal topography.
Equations (\ref{tau_zz_a})-(\ref{tau_zz_aaa}) constitute the fundamental basis for the non-hydrostatic dispersive model development.
With (\ref{tau_zz_aaa}), (\ref{tau_zz_a}) takes the form:
{\small
\\[-5mm]
\begin{linenomath*}
 \begin{eqnarray}
\begin{array}{lll}
\displaystyle{\tau_{zz_s}(\eta) = \lb 1 - \gamma^f_s\rb g^z(h-\eta)+ h\frac{\partial}{\partial t}\left [ \bar w_s
- \gamma^f_s \mathcal C^{s,f} \lb \bar w_f - \bar w_s\rb
- \gamma^{fs}_s \mathcal C^{s,fs} \lb \bar w_{fs} - \bar w_s\rb
\right ]}
\\[3mm]
\hspace{1.2cm}
\displaystyle{
+h \left [
{\bf u}_s \cdot \nabla \bar w_s 
- \gamma^f_s \mathcal C^{s,f}\lb {\bf u}_f \cdot \nabla \bar w_f - {\bf u}_s \cdot \nabla \bar w_s  \rb 
- \gamma^{fs}_s \mathcal C^{s,fs}\lb {\bf u}_{fs} \cdot \nabla \bar w_{fs} - {\bf u}_s \cdot \nabla \bar w_s  \rb
\right ] }
\\[3mm]
\hspace{1.2cm}
\displaystyle{
- \frac{\partial}{\partial t}\left [ \hat I_s 
- \gamma^f_s \mathcal C^{s,f} \lb \hat I_f - \hat I_s\rb
- \gamma^{fs}_s \mathcal C^{s,fs} \lb \hat I_{fs} - \hat I_s\rb
\right ] }\\[3mm]
\hspace{1.2cm}
\displaystyle{
- \nabla\cdot\left [ \hat I_s{\bf u}_s 
- \gamma^f_s \mathcal C^{s,f} \lb \hat I_f {\bf u}_f - \hat I_s {\bf u}_s\rb
- \gamma^{fs}_s \mathcal C^{s,fs} \lb \hat I_{fs} {\bf u}_{fs} - \hat I_s {\bf u}_s\rb
\right ]}\\[3mm]
\hspace{1.2cm}
\displaystyle{
- \left [ w_s^2 
- \gamma^f_s \mathcal C^{s,f} \lb w_f^2 - w_s^2\rb
- \gamma^{fs}_s \mathcal C^{s,fs} \lb w_{fs}^2 - w_s^2\rb
\right ]}\\[3mm] 
\hspace{1.2cm}
\displaystyle{
- \frac{1}{\alpha_s}C_{DG}^{s,f}|{\bf u}_f -{\bf u}_s|^{j-1}\left [ \lb h\bar w_f- \hat I_f\rb - \lb h\bar w_s -\hat I_s\rb\right ]}\\[3mm]
\hspace{1.2cm}
\displaystyle{
- \frac{1}{\alpha_s} C_{DG}^{s,fs}|{\bf u}_{fs} -{\bf u}_s|^{j-1}\left [ \lb h\bar w_{fs} - \hat I_{fs}\rb - \lb h \bar w_{s} -\hat I_s\rb\right ]
+C_{DV}^s\left[\left( h\bar w_s-\hat I_s\right)\right]|{\bf u}_s|.}
\label{tau_zz_aaaa}
\end{array}    
\end{eqnarray}
\end{linenomath*}
}
\hspace{-3mm}
In this representation, the first term on the right hand side contains the complementary relative flow depth, $(h-\eta)$, and indicates that at the bottom $(\eta = 0)$ it is $\lb 1 - \gamma^f_s\rb g^z h$, and at the free surface $(\eta = h)$ it is zero. So, that term is the usual hydrostatic normal load often used in shallow flow models together with the buoyancy effect $\lb 1 - \gamma^f_s\rb$. Thus, the appearance of $(h-\eta)$ in $\lb 1 - \gamma^f_s\rb g^z(h-\eta)$ implies its linear distribution from the bottom to the free surface, it is advantageous. Therefore, we should also try to transfer the other terms in (\ref{tau_zz_aaaa}) such that they contain some functions of $(h-\eta)$ and/or $\eta$. This will be achieved next.
\\[3mm]
With its definition in (\ref{tau_zz_aaa}), $\hat I_{s}$ (similar for fine-solid and fluid) can be obtained from (\ref{tau_zz_aa}) as:
\begin{linenomath*}
 \begin{eqnarray}
\begin{array}{lll}
\displaystyle{\hat I_{s} = w_{b_s} \eta - \lb \nabla\cdot {\bf u}_s\rb\frac{\eta^2}{2}}.
\label{tau_zz_aaaaa}
\end{array}    
\end{eqnarray}
\end{linenomath*}
This helps in producing desired terms with factors $h-\eta$ and/or $\eta$; see below.

\subsubsection{Effective Normal Loads}

{\bf A. The Solid Normal Load:} Now, define $D/Dt = \partial/\partial t + {\bf u}_s\cdot \nabla$ (similar for fine-solid and fluid). Then, with (\ref{tau_zz_aaaaa}), following the procedures in Yuan et al. (2018), after a lengthy calculations, (\ref{tau_zz_aaaa}) takes the form:
{\small
\begin{linenomath*}
 \begin{eqnarray}
\begin{array}{lll}
{\displaystyle
\tau_{zz_s} = \lb 1 - \gamma^f_s\rb g^z(h-\eta)
+\frac{D}{D t} \left[ \bar w_s 
-\gamma^f_s \mathcal C^{s,f}\lb \bar w_f - \bar w_s\rb  
-\gamma^{fs}_s \mathcal C^{s,fs}\lb \bar w_{fs} - \bar w_s\rb  
\right](h-\eta)}\\[3mm]
\hspace{.7cm}
{\displaystyle-\frac{1}{2}\Bigg \{\frac{D}{D t} \left[ 
h \nabla\cdot {\bf u}_s 
- \gamma^f_s \mathcal C^{s,f}\lb h \nabla\cdot {\bf u}_f - h \nabla\cdot {\bf u}_s\rb
- \gamma^{fs}_s \mathcal C^{s,fs}\lb h \nabla\cdot {\bf u}_{fs} - h \nabla\cdot {\bf u}_s\rb
\right ]\eta}\\[5mm]
\hspace{.7cm}
{\displaystyle
-\frac{D}{D t} \left[ 
 \nabla\cdot {\bf u}_s 
- \gamma^f_s \mathcal C^{s,f}\lb  \nabla\cdot {\bf u}_f -  \nabla\cdot {\bf u}_s\rb
- \gamma^{fs}_s \mathcal C^{s,fs}\lb  \nabla\cdot {\bf u}_{fs} -  \nabla\cdot {\bf u}_s\rb
\right ]\eta^2}\\[3mm]
\hspace{.7cm}
{\displaystyle
+ \left[ 
 \lb\nabla\cdot {\bf u}_s\rb^2 
- \gamma^f_s \mathcal C^{s,f}\lb  \lb\nabla\cdot {\bf u}_f\rb^2 -  \lb\nabla\cdot {\bf u}_s\rb^2\rb
- \gamma^{fs}_s \mathcal C^{s,fs}\lb \lb \nabla\cdot {\bf u}_{fs}\rb^2 -  \lb\nabla\cdot {\bf u}_s\rb^2\rb
\right ]\eta^2
\Bigg\}}
\\[3mm] 
\hspace{.7cm}
{\displaystyle
- \frac{1}{\alpha_s}C_{DG}^{s,f}|{\bf u}_f -{\bf u}_s|^{j-1}
\left [ 
\lb \bar w_f - \bar w_s\rb \lb h - \eta\rb 
- \lb \nabla\cdot\lb {\bf u}_f -{\bf u}_s\rb\rb\frac{1}{2}\eta \lb h -\eta\rb
 \right ]}\\[3mm]
 \hspace{.7cm}
 {\displaystyle
- \frac{1}{\alpha_s}C_{DG}^{s,fs}|{\bf u}_{fs} -{\bf u}_s|^{j-1}\left [ 
\lb \bar w_{fs} - \bar w_s\rb \lb h - \eta\rb 
- \lb \nabla\cdot\lb {\bf u}_{fs} -{\bf u}_s\rb\rb\frac{1}{2}\eta \lb h -\eta\rb
 \right ]}\\[3mm]
 \hspace{.7cm}
 \displaystyle{
 +\,C_{DV}^s|{\bf u}_s|\left[ \bar w_s\lb h-\eta\rb - \lb\nabla\cdot {\bf u}_s\rb\frac{1}{2}\eta\lb h-\eta\rb\right]
,}
\label{tau_zz_aaaaaa}
\end{array}    
\end{eqnarray}
\end{linenomath*}
}
\hspace{-3mm}
which is the effective normal load for the solid component. 
Note that (\ref{tau_zz_aaa}) and (\ref{tau_zz_aaaaa}) are utilized to obtain the structures associated with the drags. 
$\tau_{zz_s}$ in (\ref{tau_zz_aaaaaa}) is written entirely in terms of the flow variables, flow dynamics and the phase-interaction terms. There are two types of terms in (\ref{tau_zz_aaaaaa}). First, the slope normal acceleration terms associated with $(h-\eta)$, which are linear in $\eta$. Second, the slope parallel (divergence, or flux) terms that are either linear or quadratic in $\eta$. 
However, it is interesting to note that the interfacial drag contributions have two types of terms. First, in $C_{DG}^{s,f}$, the associated term $\lb \bar w_f - \bar w_s\rb \lb h - \eta\rb$ has a factor $\lb h - \eta\rb$ as in the usual gravity and the acceleration terms ($g^z$ and $D/Dt$). This term vanishes at the free surface. Second, $\lb \nabla\cdot\lb {\bf u}_{f} -{\bf u}_s\rb\rb\frac{1}{2}\eta \lb h -\eta\rb$ is quadratic in $\eta$, but has a special form. Such term with factor $\eta \lb h -\eta\rb$ does not appear in other contributions in $\tau_{zz_s}$. This vanishes both at the bottom and at the free surface of the flow and thus has maximum in between the flow depth. Similar analysis holds for the terms associated with $C_{DG}^{s,fs}$. So, the interfacial drags bring completely new mechanisms in the non-hydrostatic (dispersion) relations. The important point now is that, due to their structures, the first terms in the drag contributions must be (or better to) put together with the gravity and the acceleration terms, $g^z$ and $D/Dt$ (associated with $\bar w$). We consider these terms together in obtaining the enhanced gravity. Furthermore, the $D/Dt$ are due to the normal acceleration of the solid particles, and the relative acceleration of the solid particles with respect to the fine-solid and fluid. So, all $g^z, D/Dt$ and $C_{DG}$ terms (associated with $(h-\eta)$) basically represent the normal acceleration, or force. All the other remaining terms in (\ref{tau_zz_aaaaaa}) represent the dynamics and forcings in the slope parallel direction. For this reason, we re-write (\ref{tau_zz_aaaaaa}) as the first group of terms with the factor $(h-\eta)$, containing the usual gravity (including buoyancy, $\lb 1 - \gamma^f_s\rb g^z$), and the normal acceleration ($D/Dt$ terms including virtual mass) and drag terms ($C_{DG}$), and the second group of terms with $\eta$ and $\eta^2$ representing the slope parallel motion as: 
{\small
\begin{linenomath*}
 \begin{eqnarray}
\begin{array}{lll}
\displaystyle{
\tau_{zz_s} = \lb 1 - \gamma^f_s\rb g^z(h-\eta)
+\frac{D}{D t} \left[ \bar w_s 
-\gamma^f_s \mathcal C^{s,f}\lb \bar w_f - \bar w_s\rb  
-\gamma^{fs}_s \mathcal C^{s,fs}\lb \bar w_{fs} - \bar w_s\rb  
\right](h-\eta)}\\[3mm]
\hspace{.7cm}
\displaystyle{
- \frac{1}{\alpha_s}\left [C_{DG}^{s,f}|{\bf u}_f -{\bf u}_s|^{j-1}
\lb \bar w_f - \bar w_s\rb 
+ C_{DG}^{s,fs}|{\bf u}_{fs} -{\bf u}_s|^{j-1} 
\lb \bar w_{fs} - \bar w_s\rb
-C_{DV}^s\bar w_s |{\bf u}_s|\alpha_s
\right] \lb h - \eta\rb,}
\\[3mm] 
\hspace{.7cm}
\displaystyle{
-\frac{1}{2}\Bigg \{\frac{D}{D t} \left[ 
h \nabla\cdot {\bf u}_s 
- \gamma^f_s \mathcal C^{s,f}\lb h \nabla\cdot {\bf u}_f - h \nabla\cdot {\bf u}_s\rb
- \gamma^{fs}_s \mathcal C^{s,fs}\lb h \nabla\cdot {\bf u}_{fs} - h \nabla\cdot {\bf u}_s\rb
\right ]\eta}\\[3mm]
\hspace{.7cm}
\displaystyle{
-\frac{D}{D t} \left[ 
 \nabla\cdot {\bf u}_s 
- \gamma^f_s \mathcal C^{s,f}\lb  \nabla\cdot {\bf u}_f -  \nabla\cdot {\bf u}_s\rb
- \gamma^{fs}_s \mathcal C^{s,fs}\lb  \nabla\cdot {\bf u}_{fs} -  \nabla\cdot {\bf u}_s\rb
\right ]\eta^2}\\[3mm]
\hspace{.7cm}
+ \left[ 
 \lb\nabla\cdot {\bf u}_s\rb^2 
- \gamma^f_s \mathcal C^{s,f}\lb  \lb\nabla\cdot {\bf u}_f\rb^2 -  \lb\nabla\cdot {\bf u}_s\rb^2\rb
- \gamma^{fs}_s \mathcal C^{s,fs}\lb \lb \nabla\cdot {\bf u}_{fs}\rb^2 -  \lb\nabla\cdot {\bf u}_s\rb^2\rb
\right ]\eta^2
\Bigg\}
\\[3mm] 
\hspace{.7cm}
\displaystyle{
+\!\frac{1}{\alpha_s}\left[ C_{DG}^{s,f}|{\bf u}_f\! -\!{\bf u}_s|^{j-1}
\lb \nabla\cdot\lb {\bf u}_f \!-\!{\bf u}_s\rb\rb
\!+\! C_{DG}^{s,fs}|{\bf u}_{fs} \!-\!{\bf u}_s|^{j-1} \lb \nabla\cdot\lb {\bf u}_{fs} \!-\!{\bf u}_s\rb\rb
\!-\!C_{DV}^s|{\bf u}_s|\lb\nabla\cdot {\bf u}_s\rb\alpha_s 
\right ]\frac{1}{2}\eta \lb h \!-\!\eta\rb.}
\label{tau_zz_aaaaaa_group}
\end{array}    
\end{eqnarray}
\end{linenomath*}
}
\hspace{-3mm}
So, it is legitimate to call the first group of terms (with factor $h-\eta$) the enhanced gravity, and the second group of terms (with factors $\eta, \eta^2$ and $\eta (h-\eta)$) the dispersion. Together, they constitute the (effective) non-hydrostatic normal load. This has been discussed in more detail later in Section 2.1.3 and Section 2.1.4. In (\ref{tau_zz_aaaaaa_group}), the components in the drag terms have been splitted in to normal and slope parallel-type components contributing to the enhanced gravity and dispersion relations. 
\\[3mm]
To apply the normal loads in a depth-averaged formulation, we need to depth-average $\tau_{zz_s}$ in (\ref{tau_zz_aaaaaa_group}). For this, first we define the phase-divergence in slope parallel directions as: 
$\displaystyle{U_s =  \nabla \cdot {\bf u}_s,\, 
U_{fs} =  \nabla \cdot {\bf u}_{fs},\, 
U_f =  \nabla \cdot {\bf u}_f},$ then following Yuan et al. (2018), we 
 integrate (\ref{tau_zz_aaaaaa_group}) through the flow depth to obtain its mean: 
 {\small
\begin{linenomath*}
 \begin{eqnarray}
\begin{array}{lll}
\displaystyle{
\bar\tau_{zz_s} = \lb 1 - \gamma^f_s\rb g^z\frac{1}{2}h^2
+\frac{D}{D t} \left[ \bar w_s 
-\gamma^f_s \mathcal C^{s,f}\lb \bar w_f - \bar w_s\rb  
-\gamma^{fs}_s \mathcal C^{s,fs}\lb \bar w_{fs} - \bar w_s\rb  
\right]\frac{1}{2}h^2
}
\\[3mm]
\hspace{.75cm}
\displaystyle{
- \frac{1}{\alpha_s}\frac{h^2}{2}\left [C_{DG}^{s,f}|{\bf u}_f -{\bf u}_s|^{j-1} 
\lb \bar w_f - \bar w_s\rb
+C_{DG}^{s,fs}|{\bf u}_{fs} -{\bf u}_s|^{j-1} 
\lb \bar w_{fs} - \bar w_s\rb
-C_{DV}^s\bar w_s|{\bf u}_s|\alpha_s
\right ]}
\\[3mm] 
\hspace{.75cm}
\displaystyle{+\frac{h^3}{12}\left[ \lb U_s^2 
-\gamma^f_s \mathcal C^{s,f} \lb U_f^2-U_s^2\rb
-\gamma^{fs}_s \mathcal C^{s,fs} \lb U_{fs}^2-U_s^2\rb
\rb 
- \frac{D}{Dt} \lb U_s 
- \gamma^f_s\mathcal C^{s,f}\lb U_f - U_s \rb
- \gamma^{fs}_s\mathcal C^{s,fs}\lb U_{fs} - U_s \rb
\rb \right]} \\[3mm]
\hspace{.75cm}
\displaystyle{
+\frac{1}{\alpha_s}\frac{h^3}{6}\left [
C_{DG}^{s,f}|{\bf u}_f -{\bf u}_s|^{j-1} \lb U_f - U_s\rb
+C_{DG}^{s,{fs}}|{\bf u}_{fs} -{\bf u}_s|^{j-1} \lb U_{fs} - U_s\rb
-C_{DV}^s|{\bf u}_s| U_s\alpha_s
\right ],}
\label{tau_zz_aaaaaab}
\end{array}    
\end{eqnarray}
\end{linenomath*}
\small}
\hspace{-2mm}
which is the depth averaged effective solid normal load.
\\[3mm]
{\bf B. The Fine-solid and Fluid Normal Loads:} As in (\ref{w_solid_momentum}), we consider the normal components of the fine-solid and fluid momentum equations (Pudasaini, 2012; Pudasaini and Mergili, 2019). Then, following the procedure from (\ref{tau_zz_a}) to (\ref{tau_zz_aaaaaab}), we obtain the depth-averaged normal stresses for fine-solid and fluid, respectively:
\clearpage
 {\footnotesize
\begin{linenomath*}
 \begin{eqnarray}
\begin{array}{lll}
\displaystyle{
\bar\tau_{zz_{fs}} = \gamma^f_{fs} g^z\frac{1}{2}h^2
+\frac{D}{D t} \left[ \bar w_{fs} 
-\gamma^f_{fs} \mathcal C^{fs,f}\lb \bar w_f - \bar w_{fs}\rb  
+\alpha^{s}_{fs} \mathcal C^{s,fs}\lb \bar w_{fs} - \bar w_s\rb  
\right]\frac{1}{2}h^2}\\[3mm]
\hspace{.75cm}
\displaystyle{
- \frac{1}{\alpha_{fs}}\frac{h^2}{2}\left [-\frac{1}{\gamma^{fs}_s}C_{DG}^{s,fs}|{\bf u}_{fs} -{\bf u}_s|^{j-1} 
\lb \bar w_{fs} - \bar w_s\rb
+C_{DG}^{fs,f}|{\bf u}_{f} -{\bf u}_{fs}|^{j-1} 
\lb \bar w_{f} - \bar w_{fs}\rb
-C_{DV}^{fs}\bar w_{fs}|{\bf u}_{fs}|\alpha_{fs}
\right ]}
\\[3mm] 
\hspace{.75cm}
\displaystyle{+\frac{h^3}{12}\left[ \lb U_{fs}^2 
-\gamma^f_{fs} \mathcal C^{fs,f} \lb U_f^2-U_{fs}^2\rb
+\alpha^{s}_{fs} \mathcal C^{s,fs} \lb U_{fs}^2-U_s^2\rb
\rb 
- \frac{D}{Dt} \lb U_{fs} 
- \gamma^f_{fs}\mathcal C^{fs,f}\lb U_f - U_{fs} \rb
+ \alpha^{s}_{fs}\mathcal C^{s,fs}\lb U_{fs} - U_s \rb
\rb \right]} \\[3mm]
\hspace{.75cm}
\displaystyle{
+\frac{1}{\alpha_{fs}}\frac{h^3}{6}\left [-\frac{1}{\gamma^{fs}_s}
C_{DG}^{s,fs}|{\bf u}_{fs} -{\bf u}_s|^{j-1} \lb U_{fs} - U_s\rb
+C_{DG}^{fs,{f}}|{\bf u}_{f} -{\bf u}_{fs}|^{j-1} \lb U_{f} - U_{fs}\rb
-C_{DV}^{fs}|{\bf u}_{fs}| U_{fs}\alpha_{fs}
\right ],}
\label{tau_zz_aaaaaab_fs}
\end{array}    
\end{eqnarray}
\end{linenomath*}
}
{\footnotesize
\begin{linenomath*}
 \begin{eqnarray}
\begin{array}{lll}
\displaystyle{
\bar\tau_{zz_f} = g^z\frac{1}{2}h^2
+\frac{D}{D t} \left[ \bar w_f 
+\alpha^s_f \mathcal C^{s,f}\lb \bar w_f - \bar w_s\rb  
+\alpha^{fs}_f \mathcal C^{fs,f}\lb \bar w_{f} - \bar w_{fs}\rb  
\right]\frac{1}{2}h^2}\\[3mm]
\hspace{.75cm}
\displaystyle{
+ \frac{1}{\alpha_{f}}\frac{h^2}{2}\left [\frac{1}{\gamma^f_s}C_{DG}^{s,f}|{\bf u}_f -{\bf u}_s|^{j-1} 
\lb \bar w_f - \bar w_s\rb
+\frac{1}{\gamma^f_{fs}}C_{DG}^{fs,f}|{\bf u}_{f} -{\bf u}_{fs}|^{j-1} 
\lb \bar w_{f} - \bar w_{fs}\rb
+C_{DV}^{f}\bar w_{f}|{\bf u}_{f}|\alpha_{f}
\right ]}
\\[3mm] 
\hspace{.75cm}
\displaystyle{+\frac{h^3}{12}\left[ \lb U_f^2 
+\alpha^s_f \mathcal C^{s,f} \lb U_f^2-U_s^2\rb
+\alpha^{fs}_f \mathcal C^{fs,f} \lb U_{f}^2-U_{fs}^2\rb
\rb 
- \frac{D}{Dt} \lb U_f 
+ \alpha^s_f\mathcal C^{s,f}\lb U_f - U_s \rb
+ \alpha^{fs}_f\mathcal C^{fs,f}\lb U_{f} - U_{fs} \rb
\rb \right]} \\[3mm]
\hspace{.75cm}
\displaystyle{
-\frac{1}{\alpha_{f}}\frac{h^3}{6}\left [\frac{1}{\gamma^f_s}
C_{DG}^{s,f}|{\bf u}_f -{\bf u}_s|^{j-1} \lb U_f - U_s\rb
+\frac{1}{\gamma^f_{fs}}C_{DG}^{fs,{f}}|{\bf u}_{f} -{\bf u}_{fs}|^{j-1} \lb U_{f} - U_{fs}\rb
+C_{DV}^{f}|{\bf u}_{f}| U_{f}\alpha_{f}
\right ].}
\label{tau_zz_aaaaaab_f}
\end{array}    
\end{eqnarray}
\end{linenomath*}
}
\hspace{-3mm}
The first terms on the right hand sides in (\ref{tau_zz_aaaaaab})-(\ref{tau_zz_aaaaaab_f}) show the distinct scalings for the solid, fine-solid, and fluid-phases in the three-phase mixture flow. The solid and fine-solid pressures are reduced due to respective buoyancies by the factors $\lb 1- \gamma_{s}^{f}\rb$ and $\gamma_{fs}^f$. The buoyancy reduced normal load of the solid particles, $\lb 1- \gamma_{s}^{f}\rb$, is due to the {fluid} composed of water and very fine particles and the fine-solids, and thus $\gamma_{s}^{f}$ is the corresponding mixture fluid density normalized by the solid density. Similar statement holds for fine-solid. For more detail on this, see Pudasaini and Mergili (2019).
\\[3mm]
The mean values of the normal components of stresses are required to obtain the lateral (slope parallel) stress components, which for solid, fine-solid and fluid phases are given by:
$\alpha_s \bar \tau_{{xx}_s} = \alpha_s K_s^x\bar \tau_{{zz}_s}, 
\alpha_{fs} \bar \tau_{{xx}_{fs}} = \alpha_{fs}\bar \tau_{{zz}_{fs}}, 
\alpha_{f} \bar \tau_{{xx}_{f}} = \alpha_{f}\bar \tau_{{zz}_{f}},$
where only the solid-phase contains the earth pressure coefficient $K_s^x$ due to its Coulomb frictional behavior (Pudasaini, 2012; Pudasaini and Mergili, 2019). These lateral stresses enter the momentum balance equations as the sum of the enhanced hydraulic pressure gradients and dispersion relations. This is discussed later.

\subsubsection{Enhanced (Effective) Gravities}

From (\ref{tau_zz_aaaaaa}), or (\ref{tau_zz_aaaaaab}), (and similarly from (\ref{tau_zz_aaaaaab_fs}) and (\ref{tau_zz_aaaaaab_f})), we extract the enhanced (effective) gravity for solid, fine-solid and fluid components, respectively
\begin{linenomath*}
 \begin{eqnarray}
\begin{array}{lll}
\displaystyle{
{\acute g}^z_s = \lb 1- \gamma^f_s\rb{g}^z + \frac{D}{Dt} \left[ \bar w_s 
- \gamma^f_s \mathcal C^{s,f} \lb \bar w_f - \bar w_s \rb
- \gamma^{fs}_s \mathcal C^{s,fs} \lb \bar w_{fs} - \bar w_s \rb
\right]}\\[3mm]
\hspace{.55cm}
\displaystyle{
- \frac{1}{\alpha_{s}}\left [C_{DG}^{s,f}|{\bf u}_f -{\bf u}_s|^{j-1} 
\lb \bar w_f - \bar w_s\rb
+C_{DG}^{s,fs}|{\bf u}_{fs} -{\bf u}_s|^{j-1} 
\lb \bar w_{fs} - \bar w_s\rb
-C_{DV}^s\bar w_s|{\bf u}_s|\alpha_s
\right ],} 
\\[7mm]
\displaystyle{
{\acute g}^z_{fs} = \gamma^f_{fs}{g}^z + \frac{D}{Dt} \left[ \bar w_{fs} 
- \gamma^f_{fs} \mathcal C^{fs, f} \lb \bar w_{f} - \bar w_{fs} \rb
+ \alpha^s_{fs} \mathcal C^{s, fs} \lb \bar w_{fs} - \bar w_{s} \rb
\right]}\\[3mm]
\hspace{.6cm}
\displaystyle{
- \frac{1}{\alpha_{fs}}\left [-\frac{1}{\gamma^{fs}_s}C_{DG}^{s,fs}|{\bf u}_{fs} -{\bf u}_s|^{j-1} 
\lb \bar w_{fs} - \bar w_s\rb
+C_{DG}^{fs,f}|{\bf u}_{f} -{\bf u}_{fs}|^{j-1} 
\lb \bar w_{f} - \bar w_{fs}\rb
-C_{DV}^{fs}\bar w_{fs}|{\bf u}_{fs}|\alpha_{fs}
\right ]
,}\\[7mm]
\displaystyle{{\acute g}^z_f = {g}^z + \frac{D}{Dt} \left[ \bar w_f 
+ \alpha^s_f \mathcal C^{s,f} \lb \bar w_f - \bar w_s \rb
+ \alpha^{fs}_f \mathcal C^{fs,f} \lb \bar w_f - \bar w_{fs} \rb
\right]}\\[3mm]
\hspace{.55cm}
\displaystyle{
+ \frac{1}{\alpha_{f}}\left [\frac{1}{\gamma^f_s}C_{DG}^{s,f}|{\bf u}_f -{\bf u}_s|^{j-1} 
\lb \bar w_f - \bar w_s\rb
+\frac{1}{\gamma^f_{fs}}C_{DG}^{fs,f}|{\bf u}_{f} -{\bf u}_{fs}|^{j-1} 
\lb \bar w_{f} - \bar w_{fs}\rb
+C_{DV}^f\bar w_f|{\bf u}_f|\alpha_f
\right ],}
\label{effective_gravity}
\end{array}    
\end{eqnarray}
\end{linenomath*}
where the factors $h^2/2$ do not appear due to the definition of acceleration.
These expressions can be obtained directly from (\ref{tau_zz_aaaaaa_group}) by setting $\eta \to 0$, i.e., the normal loads at the bed. This clearly indicates which terms in (\ref{tau_zz_aaaaaa_group}) contribute to the enhanced gravity or the effective normal load at the bed, and which other terms contribute to dispersive effects. For vanishing fine-solid and fluid components, these reduce to the simple enhanced gravity in Denlinger and Iverson (2004), Castro-Orgaz et al. (2015) and Yuan et al. (2018) for single-phase granular flow equations. Our new multi-phase formulations include buoyancy reduced solid and fine-solid normal loads as indicated by the factors $\lb 1- \gamma^f_s\rb$ and $\gamma^f_{fs}$, and the virtual mass forces as indicated by $\mathcal C$. The virtual mass forces alter the solid, fine-solid and fluid accelerations in the flow normal direction (in $D/Dt$) that ultimately enhance the effective gravity of the solid, fine-solid, and fluid phases. Furthermore, the drags between the phases $\lb C_{DG} \rb$ and the viscous drags $\lb C_{DV} \rb$ appear only in our enhanced gravity. Depending on the values of $\gamma, \mathcal C, C_{DG}, C_{DV}$ and the relative phase-velocities in the flow depth direction, enhancements or reductions of the usual gravity loads can be substantial to dominant as compared to the usual gravity loads, $g^z$.   
\\[3mm]
These enhanced gravity terms include the accelerations of the solid, fine-solid and fluid components in the slope normal direction indicated by $D/Dt$. Furthermore, (\ref{effective_gravity}) also includes the drag contributions in the slope normal direction.
The only common quantity in (\ref{effective_gravity}), is the usual gravity load, $g^z$. However, the enhanced gravities differ with the surface normal accelerations of the solid, fine-solid and fluid phases. Depending on the flow dynamics, interfacial momentum exchanges, viscous drags, and the boundary conditions, one or two of them could be substantially larger than the others. One prominent example is a landslide impacting a reservoir or a water body (Pudasaini, 2014; Kafle et al., 2019; Mergili et al., 2018, 2020b; Pudasaini and Mergili, 2019). In this situation, both the enhanced gravity and the dispersion (see below) of the water wave would be fundamentally different (can also be large) as compared to the enhanced gravity and the dispersion of the submarine landslide. As we will see later (\ref{effective_gravity}) are components of the full non-hydrostatic model formulation.

\subsubsection{Dispersive Contributions} 

The main dispersive contributions for the solid, fine-solid and fluid are denoted by $\mathcal D_s^x, \mathcal D_{fs}^x, \mathcal D_f^x$ which are extracted from (\ref{tau_zz_aaaaaab})-(\ref{tau_zz_aaaaaab_f}). We call them dispersive (for simplicity of terminology, also, see Castro-Orgaz et al., 2015) and take the form:
{\footnotesize
\begin{linenomath*}
\begin{eqnarray}
\begin{array}{lll}
\displaystyle{\mathcal D_s^x = K_s^x\frac{h^2}{12}\left[ \lb U_s^2 
-\gamma^f_s \mathcal C^{s,f} \lb U_f^2-U_s^2\rb
-\gamma^{fs}_s \mathcal C^{s,fs} \lb U_{fs}^2-U_s^2\rb
\rb 
- \frac{D}{Dt} \lb U_s 
- \gamma^f_s\mathcal C^{s,f}\lb U_f - U_s \rb
- \gamma^{fs}_s\mathcal C^{s,fs}\lb U_{fs} - U_s \rb
\rb \right]} \\[3mm]
\hspace{.55cm}
\displaystyle{
+\, K_s^x\frac{1}{\alpha_{s}}\frac{h^2}{6}\left [
C_{DG}^{s,f}|{\bf u}_f -{\bf u}_s|^{j-1} \lb U_f - U_s\rb
+C_{DG}^{s,{fs}}|{\bf u}_{fs} -{\bf u}_s|^{j-1} \lb U_{fs} - U_s\rb
-C_{DV}^s|{\bf u}_s|U_s\,\alpha_s
\right ],}
\\[7mm]
{\mathcal D_{fs}^x = \frac{h^2}{12}\left[ \lb U_{fs}^2 
-\gamma^f_{fs} \mathcal C^{fs,f} \lb U_f^2-U_{fs}^2\rb
+\alpha^{s}_{fs} \mathcal C^{s,fs} \lb U_{fs}^2-U_s^2\rb
\rb 
- \frac{D}{Dt} \lb U_{fs} 
- \gamma^f_{fs}\mathcal C^{fs,f}\lb U_f - U_{fs} \rb
+ \alpha^{s}_{fs}\mathcal C^{s,fs}\lb U_{fs} - U_s \rb
\rb \right]}, \\[3mm]
\hspace{.65cm}
+\frac{1}{\alpha_{fs}}\frac{h^2}{6}\left [-\frac{1}{\gamma^{fs}_s}
C_{DG}^{s,fs}|{\bf u}_{fs} -{\bf u}_s|^{j-1} \lb U_{fs} - U_s\rb
+C_{DG}^{fs,{f}}|{\bf u}_{f} -{\bf u}_{fs}|^{j-1} \lb U_{f} - U_{fs}\rb
-C_{DV}^{fs}|{\bf u}_{fs}|U_{fs}\,\alpha_{fs}
\right ],
\\[7mm]
\displaystyle{\mathcal D_f^x = \frac{h^2}{12}\left[ \lb U_f^2 
+\alpha^s_f \mathcal C^{s,f} \lb U_f^2-U_s^2\rb
+\alpha^{fs}_f \mathcal C^{fs,f} \lb U_f^2-U_{fs}^2\rb
\rb 
- \frac{D}{Dt} \lb U_f + \alpha^s_f\mathcal C^{s,f}\lb U_f - U_s \rb
+ \alpha^{fs}_f\mathcal C^{fs,f}\lb U_f - U_{fs} \rb
\rb \right]}\\[3mm]
\hspace{.6cm}
-\frac{1}{\alpha_{f}}\frac{h^2}{6}\left [\frac{1}{\gamma^f_s}
C_{DG}^{s,f}|{\bf u}_f -{\bf u}_s|^{j-1} \lb U_f - U_s\rb
+\frac{1}{\gamma^f_{fs}}C_{DG}^{fs,{f}}|{\bf u}_{f} -{\bf u}_{fs}|^{j-1} \lb U_{f} - U_{fs}\rb
+C_{DV}^f|{\bf u}_f|U_f\,\alpha_f
\right ].
\label{dispersion}
\end{array}    
\end{eqnarray}
\end{linenomath*}
}
\hspace{-3mm}
So, (\ref{effective_gravity}) and (\ref{dispersion}) imply that $\bar \tau_{zz_s} = \acute g^z_s h^2/2 + \mathcal D^x_s h/K_s^x$, 
$\bar \tau_{zz_{fs}} = \acute g^z_{fs} h^2/2 + \mathcal D^x_{fs} h$,
$\bar \tau_{zz_{f}} = \acute g^z_{f} h^2/2 + \mathcal D^x_{f} h$. Hence, the effective basal normal load is the sum of the effective gravity and (effective) dispersion. Note that the factor $h$ is taken out from the dispersion expressions to properly adjust the fluxes, because 
$\alpha_s \bar\tau_{xx_s} = \alpha_s K_s^x\acute g^z_s h^2/2 + \alpha_s \mathcal D_s^x h
= \alpha_s h \left [ \acute \beta^x_s h/2 + \mathcal D_s^x \right ]$, etc., where $\acute \beta^x_s = K_s^x\acute g^z_s$. Later, such structures will appear in the lateral fluxes in the momentum balance equations, where $\acute \beta^x_s h/2$ and $ D_s^x$ correspond, respectively, to the enhanced hydraulic pressure gradient and dispersion. 
\\[3mm]
In what follows, all the terms with ${\acute *}$ are the enhanced terms, while these and all the $\mathcal D$ terms are entirely new contributions to the Pudasaini and Mergili (2019) model. These reduce to the non-hydrostatic relations for single-phase granular flow in Denlinger and Iverson (2004), Castro-Orgaz (2015), and Yuan et al. (2018). It is important to note that the enhanced gravities (\ref{effective_gravity}) and the dispersion relations (\ref{dispersion}) are derived from the $w$ components of the momentum balances from the multi-phase phase mass flow model (Pudasaini and Mergili, 2019). So, there are direct and strong couplings between the solid, fine-solid and the fluid components among these dispersion relations. As in the effective gravity, the dispersive terms are strongly coupled, e.g., due to the interfacial drag and virtual mass contributions.

\subsection{The Non-Hydrostatic Multi-Phase Mass-Flow Model}

In what follows, we further develop the three-phase mass flow model (Pudasaini and Mergili, 2019) by incorporating the enhanced gravities (\ref{effective_gravity}) and the dispersion relations (\ref{dispersion}). The depth-averaged mass balance equations for the solid, fine-solid and fluid phases are:
 \begin{linenomath*}
\begin{subequations}\label{Model_Final_Mass}
\begin{align}
&\displaystyle{\frac{\partial }{\partial t}{\lb \alpha_s h\rb} + \frac{\partial}{\partial x}{\lb \alpha_s h u_s\rb}
                 + \frac{\partial}{\partial y}{\lb \alpha_s h v_s\rb}=0},\label{Model_Final_Mass_s}\\[1mm]
&\displaystyle{\frac{\partial }{\partial t}{\lb \alpha_{fs} h\rb} + \frac{\partial}{\partial x}{\lb \alpha_{fs} h u_{fs}\rb}
                 + \frac{\partial}{\partial y}{\lb \alpha_{fs} h v_{fs}\rb}=0},\label{Model_Final_Mass_fs}\\[1mm]                 
&\displaystyle{\frac{\partial }{\partial t}{\lb \alpha_f h\rb} + \frac{\partial}{\partial x}{\lb \alpha_f h u_f\rb}
                 + \frac{\partial}{\partial y}{\lb \alpha_f h v_f\rb}=0.\label{Model_Final_Mass_f}}
\end{align}
\end{subequations}
\end{linenomath*}
The $x$-directional depth-averaged momentum conservation equations for the solid, fine-solid and fluid phases are,
{\small
\begin{linenomath*}
\begin{subequations}\label{Model_Final_Momentum}
\begin{align}
&\hspace{-1.5cm}\displaystyle{\frac{\partial }{\partial t}\biggl [ \alpha_s h \lb u_s - u_{s}^{vm} \rb \biggr ]
  +\frac{\partial }{\partial x}\biggl [ \alpha_s h \lb u_s^2 - uu_{s}^{vm}+ {\acute \beta}_{s}^x \frac{h}{2}  + \mathcal D_s^x\rb \biggr ]
  +\frac{\partial }{\partial y}\biggl[ \alpha_s h \lb u_sv_s - uv_{s}^{vm}\rb \biggr ]}
\displaystyle{=  h\mathcal S_{s}^x}, \label{Model_Final_Momentum_s}\\[5mm] 
&\hspace{-1.5cm}\displaystyle{\frac{\partial }{\partial t}\biggl [ \alpha_{fs} h \lb u_{fs} - u_{fs}^{vm} \rb \biggr ]
  +\frac{\partial }{\partial x}\biggl [ \alpha_{fs} h \lb u_{fs}^2 - uu_{fs}^{vm}+ {\acute\beta}_{fs}^x \frac{h}{2} + \mathcal D_{fs}^x \rb \biggr ]
  +\frac{\partial }{\partial y}\biggl[ \alpha_{fs} h \lb u_{fs}v_{fs} - uv_{fs}^{vm}\rb \biggr ]}
\displaystyle{=  h\mathcal S_{fs}^x},\label{Model_Final_Momentum_fs}\\[5mm] 
&\hspace{-1.5cm}\displaystyle{\frac{\partial }{\partial t}\biggl [ \alpha_f h \lb u_f + u_{f}^{vm} \rb \biggr ]
  +\frac{\partial }{\partial x}\biggl [ \alpha_f h \lb u_f^2 + uu_{f}^{vm}+ {\acute\beta}_{f}^x \frac{h}{2} + \mathcal D_f^x\rb \biggr ]
  +\frac{\partial }{\partial y}\biggl[ \alpha_f h \lb u_fv_f + uv_{f}^{vm}\rb \biggr ]}
\displaystyle{=  h\mathcal S_{f}^x}.\label{Model_Final_Momentum_f}%\\[5mm] 
\end{align}
\end{subequations}
\end{linenomath*}
}
\hspace{-3mm}
It is evident that the enhancements of the momentum fluxes depend on ${\acute\beta}$ and $\mathcal D$. Since the flow depth $h$ is a common factor in the momentum fluxes, the terms associated with $\mathcal D$ are proportional to $h^3$, and the term associated with ${\acute\beta}$ are proportional to $h^2$. This, together with the structure of $\mathcal D$ and ${\acute\beta}$ in the fluxes in (\ref{Model_Final_Momentum}), signify the highly non-linear, non-hydrostatic (dispersion) contributions.
\\[3mm]
Due to symmetry, the $y$-directional momentum equations for the solid, fine-solid and fluid phases can be written similarly {here and in all the following considerations}. {This is achieved by formally utilizing the replacements: $x \longleftrightarrow y$ and $u \longleftrightarrow v$, whenever necessary, both for variables and associated parameters.}
Below, we present models for all the fluxes, and source terms {and forces in the momentum equations for multi-phase mass flows where we follow the structures in Pudasaini and Mergili (2019).} First, we write those terms that include the non-hydrostatic terms (enhanced gravity and dispersion). The other terms are as in Pudasaini and Mergili (2019) and are put in an Appendix for completeness. It is important to note that in structure (\ref{Model_Final_Mass})-(\ref{Model_Final_Momentum}) are the same as in Pudasaini and Mergili (2019). It is advantageous, because the similar analysis and numerical methods and tools as in Pudasaini and Mergili (2019) might be applied to solve the new system of non-hydrostatic multi-phase mass flow model. However, complexity arises due to the new non-hydrostatic terms, particularly associated with the higher order time and spatial derivatives.
\\[1mm]
{\bf The $x$-directional source terms} in (\ref{Model_Final_Momentum}) are
\begin{linenomath*}
\begin{subequations}\label{Source}
\begin{align}
\mathcal S_{s}^x &= \alpha_s\left [g^x - \frac{u_s}{|{\bf u}_s|}\tan\delta_s {\acute g}^z_s -{\acute g}^z_s\frac{\partial b}{\partial x}\right ] - \alpha_s {{g''}}^z_s \left [ \frac{\partial h}{\partial x} + \frac{\partial b}{\partial x}\right ]\nonumber\\[2mm]
&+ C_{DG}^{s,f}  \lb u_f - u_s \rb{ |{\bf u}_f - {\bf u}_s|}^{\jmath-1}
 + C_{DG}^{s,fs} \lb u_{fs} - u_s \rb{ |{\bf u}_{fs} - {\bf u}_s|}^{\jmath-1}
 -C_{DV}^s u_s |{\bf u}_s|\alpha_s,\label{Source_x_s}\\[2mm]
\mathcal S_{fs}^x &= \alpha_{fs}\biggl [g^x - \biggl [ - \frac{1}{2} {\acute g}^z_{fs}  \frac{h}{\alpha_{fs}}\frac{\partial \alpha_{fs}}{\partial x} +  {\acute g}^z_{fs} \frac{\partial b}{\partial x}\nonumber \\[2mm]
& -\left \{ 2\frac{\partial}{\partial x}\lb \nu_{fs}^e\frac{\partial u_{fs}}{\partial x}\rb
+  \frac{\partial}{\partial y}\lb \nu_{fs}^e\frac{\partial v_{fs}}{\partial x}\rb
 + \frac{\partial}{\partial y}\lb \nu_{fs}^e\frac{\partial u_{fs}}{\partial y}\rb
 -  \nu_{fs}^e\left [ \frac{\partial u_{fs}}{\partial z}\right ]_b\frac{1}{h} \right \} + \tau_{nN}^{fs^x} \biggr]\biggl ]\nonumber \\[2mm]
&-\displaystyle\frac{1}{\gamma_s^{fs}}{C_{DG}^{s,fs}\lb u_{fs} - u_s \rb{ |{\bf u}_{fs} - {\bf u}_s|}^{\jmath-1}}
 +\displaystyle{C_{DG}^{fs,f}\lb u_f - u_{fs} \rb{ |{\bf u}_f - {\bf u}_{fs}|}^{\jmath-1}}
 -C_{DV}^{fs} u_{fs} |{\bf u}_{fs}|\alpha_{fs},\label{Source_x_fs}
\\[2mm]
\mathcal S_{f}^x &= \alpha_f\biggl [g^x - \biggl [ - \frac{1}{2} {\acute g}^z_{f}\frac{h}{\alpha_f}\frac{\partial \alpha_f}{\partial x} +  {\acute g}^z_{f}\frac{\partial b}{\partial x}\nonumber \\[2mm]
& -\left \{ 2\frac{\partial}{\partial x}\lb \nu_f^e\frac{\partial u_f}{\partial x}\rb
+  \frac{\partial}{\partial y}\lb \nu_f^e\frac{\partial v_f}{\partial x}\rb
 + \frac{\partial}{\partial y}\lb \nu_f^e\frac{\partial u_f}{\partial y}\rb
 - \nu_{f}^e\left [ \frac{\partial u_f}{\partial z}\right ]_b\frac{1}{h} \right \}  + \tau_{nN}^{f^x} \biggr]\biggl ]\nonumber \\[2mm]
&-\displaystyle{\frac{1}{\gamma_s^f}C_{DG}^{s,f}\lb u_f - u_s \rb{ |{\bf u}_f - {\bf u}_s|}^{\jmath-1}}
 -\displaystyle{\frac{1}{\gamma_{fs}^f}C_{DG}^{fs,f}\lb u_f - u_{fs} \rb{ |{\bf u}_f - {\bf u}_{fs}|}^{\jmath-1}}
 -C_{DV}^f u_f |{\bf u}_f|\alpha_f,\label{Source_x_f}
\end{align}
\end{subequations} 
\end{linenomath*}
where ${{g''}}^z_s$ is obtained from ${\acute g}^z_s$ by replacing $\lb 1- \gamma^f_s\rb g^z$ by $\gamma^f_s g^z$ while the other terms remain unchanged. The expressions in (\ref{Source}) are more general than those in Pudasaini and Mergili (2019) as they include the non-hydrostatic effects together with the interfacial momentum transfers. The structure of ${\acute g}$ indicates that the enhancements of the forces associated with ${\acute g}$, including friction, buoyancy and basal and topographic pressure gradients, depend on the sign and magnitude of ${\acute g}$.
\\[3mm]
Due to the acceleration in the slope normal direction, in (\ref{Source}), the solid velocity is given by ${\bf u}_s = \lb u_s, v_s, w_s\rb$, where $w_s = u_s\partial b/\partial x + v_s\partial b/\partial y$ (Yuan et al., 2018). This indicates that for locally changing basal topography, the surface normal component of velocity is important. Similar expressions hold for the fine-solid and fluid components. 
\\[3mm]
In (\ref{Model_Final_Momentum}) and (\ref{Source}), 
$u^{vm}, uu^{vm}, uv^{vm}$ are the virtual mass induced mass and momentum enhancements, 
${\acute \beta}$ are the hydraulic pressure coefficients, 
$\nu^e$ are the effective kinematic viscosities, $\partial u/\partial z|_b$ are the $xz$- basal shear stresses, $\tau_{nN}$ are the enhanced non-Newtonian viscous stresses, and 
$C_{DG}$ are the drag coefficients. 
The momentum balances (\ref{Model_Final_Momentum}) and the sources (\ref{Source}) indicate that the effective gravity enhances the ``hydraulic pressure gradients'' (via the terms associated with ${\acute \beta}$) in the momentum flux, and the enhanced material loads at the base as indicated by the terms associated with ${\acute g}$ and $g''$, indicating their extensive effects in the source terms. In total, the lateral flux for solid is enhanced by 
$\alpha_s \left [ {\acute \beta}_s^x - \beta_s^x\right ]h^2/2 +\alpha_s h {\mathcal D}_s^x$, where, $\beta_s^x = K_s^x\lb 1- \gamma_s^f \rb g^z$.
Similar flux enhancements emerge for the fine-solid and fluid phases.
\\[3mm]
{\bf The $x$-directional hydraulic pressure coefficients} for solid, fine-solid and fluid in (\ref{Model_Final_Momentum}) are:
\begin{linenomath*}
\begin{eqnarray}
\begin{array}{lll}\label{hydraulic_pressure_x}
{\acute\beta}_{s}^x = K_{s}^x  {\acute g}^z_s,\,\,\, {\acute\beta}_{fs}^x = {\acute g}^z_{fs},\,\,\,{\acute\beta}_{f}^x = {\acute g}^z_f,
\label{pressure_parameter_scaling}
\end{array}    
\end{eqnarray}
\end{linenomath*}
where {$K_s^x$ is the earth pressure coefficient (Pudasaini and Hutter, 2007)} and ${\acute g}$ are given by (\ref{effective_gravity}). Above, we only wrote those terms that are new in the non-hydrostatic formulations, that are ${\acute \beta}, {\acute g}, g''$ and $\mathcal D$. Based on Pudasaini and Mergili (2019), all other terms appearing in the above model equations are explained in the Appendix.
\\[3mm]
{\bf A Closed System of Equations:} The model (\ref{Model_Final_Mass})-(\ref{Model_Final_Momentum}) constitutes a set of nine equations for mass and momentum balances (including the $y$-components) for three-phase mixture mass flows in nine unknowns, namely, the solid, fine-solid and fluid phase velocities in the down-slope $\lb u_s, u_{fs}, u_f\rb$, and cross slope $\lb v_s, v_{fs}, v_f\rb$ directions, and the respective phase depths $\lb h_s=\alpha_sh, h_{fs}=\alpha_{fs}h, h_f=\alpha_fh \rb$.
{Note that $h_s + h_{fs} + h_f = h$, the total material depth.}
The model is written in a well structured form of partial differential equations and may be solved numerically once appropriate initial and boundary conditions are prescribed (Pudasaini and Mergili, 2019). 
\\[3mm]
{\bf Reduction to Existing Models:} 
By setting the fine-solid and fluid fractions to zero ($\alpha_{fs} \to 0, \alpha_{f} \to 0$), the new non-hydrostatic multi-phase mass flow model reduces to the single-phase non-hydrostatic granular flow models by Castro-Orgaz et al. (2015) and Yuan et al. (2018). The major parts of $\acute g, g'', \acute \beta$ terms, and entirely the $\mathcal D$ terms in (\ref{Model_Final_Momentum})-(\ref{Source}) are new to Pudasaini and Mergili (2019) which are due to non-hydrostatic contributions. Furthermore, the Pudasaini and Mergili (2019) multi-phase mass flow model is obtained by neglecting all the non-hydrostatic contributions, i.e., by only considering
$\acute g_s^z: = \lb 1 - \gamma^f_s\rb g^z$, $\acute g_{fs}^z: = \gamma^f_{fs}g^z$, $\acute g_{f}^z: = g^z$; $\mathcal D_s^x = 0, \mathcal D_{fs}^x = 0, \mathcal D_{f}^x = 0$.

\section{Possible Simplifications}

As we saw in Section 2.2, relations (\ref{effective_gravity}) and (\ref{dispersion}) introduce higher order spatial and time derivatives in the momentum fluxes. The new enhanced gravity and dispersion may lead to a complexity in numerical integration of the model equations, and thus may require a fundamentally new and complicated numerical method to properly solve the model equations. 
That was the case even for the simple single-phase granular flow models (Castro-Orgaz et al., 2015; Yuan et al., 2018). 

\subsection{Reduced Normal Load - Ignoring the Time Derivatives in Dispersion}

One way to avoid computational difficulties, but still include the new effects, is to assume a negligible local time derivatives ($\partial/\partial t$) in (\ref{tau_zz_aaaaaab}). This can be a reasonable assumption, e.g., after the initial impact of the landslide at the water body and during continues impact. Another possibility is to ignore all the $D/Dt$ terms in (\ref{tau_zz_aaaaaab}). Yet, the reduced solid normal stress includes non-hydrostatic effects due to buoyancy, virtual mass, drags and slope parallel divergence and relative divergence,
{\small
\begin{linenomath*}
 \begin{eqnarray}
\begin{array}{lll}
\bar\tau_{zz_{sR}} = \lb 1 - \gamma^f_s\rb g^z\frac{1}{2}h^2
- \frac{1}{2}\frac{1}{\alpha_{s}}h^2\left [C_{DG}^{s,f}|{\bf u}_f -{\bf u}_s|^{j-1} 
\lb \bar w_f - \bar w_s\rb
+C_{DG}^{s,fs}|{\bf u}_{fs} -{\bf u}_s|^{j-1} 
\lb \bar w_{fs} - \bar w_s\rb
-C_{DV}^s\bar w_s |{\bf u}_s| \alpha_s
\right ]
\\[3mm] 
\hspace{.9cm}
\displaystyle{+
\frac{h^3}{12}\left[ \lb U_s^2 
-\gamma^f_s \mathcal C^{s,f} \lb U_f^2-U_s^2\rb
-\gamma^{fs}_s \mathcal C^{s,fs} \lb U_{fs}^2-U_s^2\rb
\rb 
\right]} \\[3mm]
\hspace{.9cm}
\displaystyle{
+\frac{1}{\alpha_{s}}\frac{h^3}{6}\left [
C_{DG}^{s,f}|{\bf u}_f -{\bf u}_s|^{j-1} \lb U_f - U_s\rb
+C_{DG}^{s,{fs}}|{\bf u}_{fs} -{\bf u}_s|^{j-1} \lb U_{fs} - U_s\rb
-C_{DV}^s |{\bf u}_s| U_s\,\alpha_s
\right ],}
\label{tau_zz_aaaaaab_reduced}
\end{array}    
\end{eqnarray}
\end{linenomath*}
}
\hspace{-3mm}
where $R$ in $\bar\tau_{zz_{sR}}$ stands for the reduced normal stress. And thus, the corresponding reduced enhanced gravity and reduced dispersion expressions are given, respectively, by
{\small
\begin{linenomath*}
 \begin{eqnarray}
\begin{array}{lll}
{\acute g}^z_{sR} = \lb 1 - \gamma^f_s\rb g^z
- \frac{1}{\alpha_{s}}\left [C_{DG}^{s,f}|{\bf u}_f -{\bf u}_s|^{j-1} 
\lb \bar w_f - \bar w_s\rb
+C_{DG}^{s,fs}|{\bf u}_{fs} -{\bf u}_s|^{j-1} 
\lb \bar w_{fs} - \bar w_s\rb
-C_{DV}^s\bar w_s |{\bf u}_s| \alpha_s
\right ],
\\[5mm] 
\displaystyle{
\mathcal D_{sR}^s = K_s^x\frac{h^2}{12}\left[ \lb U_s^2 
-\gamma^f_s \mathcal C^{s,f} \lb U_f^2-U_s^2\rb
-\gamma^{fs}_s \mathcal C^{s,fs} \lb U_{fs}^2-U_s^2\rb
\rb 
\right]} \\[3mm]
\hspace{.75cm}
{\displaystyle
+\frac{1}{\alpha_{s}}K_s^x\frac{h^2}{6}\left [
C_{DG}^{s,f}|{\bf u}_f -{\bf u}_s|^{j-1} \lb U_f - U_s\rb
+C_{DG}^{s,{fs}}|{\bf u}_{fs} -{\bf u}_s|^{j-1} \lb U_{fs} - U_s\rb
-C_{DV}^s |{\bf u}_s| U_s\,\alpha_s
\right ].}
\label{tau_zz_aaaaaab_reduce_g_d}
\end{array}    
\end{eqnarray}
\end{linenomath*}
}
\hspace{-3mm}
From (\ref{effective_gravity}) and (\ref{dispersion}), similar reduced expressions can be obtained for the fine-solid and fluid components. For single-phase granular flow without the fine-solid and fluid components, (\ref{tau_zz_aaaaaab_reduce_g_d}) would further drastically reduce to 
${\acute g}^z_{sR} = g^z + C_{DV}^s\bar w_s |{\bf u}_s|$ and
$\mathcal D_{sR}^x = K_s^xh^2 U_s^2/12 - \frac{1}{6}K_s^xh^2 C_{DV}^s |{\bf u}_s| U_s$. However, in general, as in (\ref{effective_gravity}) and (\ref{dispersion}), the full descriptions of ${\acute g}^z_{s}$ and $\mathcal D_{s}^x$ (similar for fine-solid and fluid components) should be considered in simulating non-hydrostatic mixture flows.

\subsection{Approximations to Time Derivatives in Dispersion and Enhanced Gravity}

One of the major difficulties associated with the non-hydrostatic model presented above is the presence of the time derivatives in enhanced gravity and dispersion. In the simple situation without interfacial drag and virtual mass, the dispersion in (\ref{Model_Final_Momentum_s}) is given by
{\footnotesize
\begin{linenomath*}
\begin{eqnarray}
\begin{array}{lll}
\displaystyle{
\frac{\partial}{\partial x}\left [\frac{K_s^x}{12}\alpha_sh^3
\left\{ U_s^2 - \frac{D U_s}{Dt} 
-2 C_{DV}^s|{\bf u}_s| U_s
\right\}
\right] =
\frac{\partial}{\partial x}\left [\frac{K_s^x}{12}\alpha_sh^3
\left\{ U_s^2 - \lb \frac{\partial}{\partial t} + u_s\frac{\partial}{\partial x}
+ v_s\frac{\partial}{\partial y}
\rb U_s 
-2 C_{DV}^s|{\bf u}_s| U_s
\right\}
\right]}\\[5mm]
= \displaystyle{
\frac{\partial}{\partial x}\left [\frac{K_s^x}{12}\alpha_sh^3
\left\{ U_s^2 - 
\lb \frac{\partial}{\partial x}\frac{\partial u_s}{\partial t}
+   \frac{\partial}{\partial y}\frac{\partial v_s}{\partial t}
\rb
-\lb u_s\frac{\partial}{\partial x} + v_s\frac{\partial}{\partial y}\rb U_s
-2 C_{DV}^s|{\bf u}_s| U_s
\right\}
\right]}.
\label{Dispersion_T}
\end{array}    
\end{eqnarray}
\end{linenomath*}
}
\hspace{-3mm}
From a computational point of view $\partial u_s/\partial t$ and $\partial v_s/\partial t$ in (\ref{Dispersion_T}) may  pose great difficulties. So, it is desirable to find expressions for $\partial u_s/\partial t$ and $\partial v_s/\partial t$ in terms of spatial derivatives, flow variables, and parameters, but no direct involvement of (the time and) time derivatives. This is a challenging task. However, we can develop simplified expressions for these for non-inertial flows. This can be achieved, e.g., by combining the simple mass and momentum balance equation for solid from (\ref{Model_Final_Mass_s}) and (\ref{Model_Final_Momentum_s}), by ignoring all extra forces (which, however, could be considered to include more complex situations). Which is equivalent to assume that all the applied forces balance each other. This results in a simple expression as:
\begin{linenomath*}
\begin{eqnarray}
\begin{array}{lll}
\displaystyle{
\frac{\partial u_s}{\partial t} =  - u_s\frac{\partial u_s}{\partial x} - v_s\frac{\partial u_s}{\partial y}.
}
\label{Dispersion_TT}
\end{array}    
\end{eqnarray}
\end{linenomath*}
Inserting (\ref{Dispersion_TT}) in to (\ref{Dispersion_T}), we technically remove $\partial u_s/\partial t$, which, however, is highly non-linear and very complex as it involves the fifth order terms (combining flow depth and velocities) and third order derivatives. Simplified expressions for the fine-solid and fluid components can be developed, and respectively take the form:
\begin{linenomath*}
\begin{eqnarray}
\begin{array}{lll}
\displaystyle{
\frac{\partial u_{fs}}{\partial t} = 
- u_{fs}\frac{\partial u_{fs}}{\partial x}
- v_{fs}\frac{\partial u_{fs}}{\partial y},
}
\label{Dispersion_TTT_fs}
\end{array}    
\end{eqnarray}
\end{linenomath*}
\begin{linenomath*}
\begin{eqnarray}
\begin{array}{lll}
\displaystyle{
\frac{\partial u_{f}}{\partial t} = 
- u_{f}\frac{\partial u_{f}}{\partial x}
- v_{f}\frac{\partial u_{f}}{\partial y}.
}
\label{Dispersion_TTT_f}
\end{array}    
\end{eqnarray}
\end{linenomath*}
Similar expressions hold for $\partial v_s/\partial t, \partial v_{fs}/\partial t$ and $\partial v_f/\partial t$. 
 Then, the dispersion term containing the time derivatives, together with $U^2$ and the viscous drag in (\ref{Dispersion_T}), reduces, for solid-phase, to:
 \begin{linenomath*}
\begin{eqnarray}
\begin{array}{lll}
\displaystyle{
U^2_s - \frac{DU_s}{Dt} 
-2 C_{DV}^s|{\bf u}_s| U_s
= 2U_s^2 
-2\frac{\partial u_s}{\partial x}\frac{\partial v_s}{\partial y}
+2\frac{\partial v_s}{\partial x}\frac{\partial u_s}{\partial y}
  -2 C_{DV}^s|{\bf u}_s| U_s      }.
\label{Dispersion_TTT_ss}
\end{array}    
\end{eqnarray}
\end{linenomath*}
Expressions for $U_{fs}^2-{DU_{fs}}/{Dt}$ and $U_{f}^2-{DU_{f}}/{Dt}$ take analogous forms. 
\\[3mm]
Similarly, with somewhat lengthy calculations, we can write the time derivative term, $D {\bar w}_s/Dt$, in the enhanced gravity (see, Section 2.1.3) as
{\small
\begin{linenomath*}
\begin{eqnarray}
\begin{array}{lll}
\displaystyle{
\frac{D {\bar w}_s}{Dt}=}
\displaystyle{ -\frac{1}{2}\left [
-h\left\{\lb \frac{\partial u_s}{\partial x}\rb^2 + 2\frac{\partial v_s}{\partial x} \frac{\partial u_s}{\partial y} + \lb \frac{\partial v_s}{\partial y}\rb^2 \right\}
+\frac{\partial h}{\partial t}\lb \frac{\partial u_s}{\partial x}+\frac{\partial v_s}{\partial y}\rb
+
\lb 
u_s\frac{\partial h}{\partial x}+v_s\frac{\partial h}{\partial y}\rb\lb\frac{\partial u_s}{\partial x}+\frac{\partial v_s}{\partial y}
\rb
\right ],
}
\label{Dispersion_TTT_ssn}
\end{array}    
\end{eqnarray}
\end{linenomath*}
}
\hspace{-3mm}
where the topographic slope changes $(\partial b/\partial x, \partial b/\partial y)$ has been ignored, which could easily be included. Similar expressions as (\ref{Dispersion_TTT_ssn}) hold for fine-solid and fluid components, ${D {\bar w}_{fs}}/{Dt}$, ${D {\bar w}_{f}}/{Dt}$. 
\\[3mm]
Due to the definition of ${\bar w}_s$, the time derivative of the flow depth, $\partial h/\partial t$, still remains in (\ref{Dispersion_TTT_ssn}). However, this can be obtained by summing-up the mass balance equations (\ref{Model_Final_Mass}) for the solid, fine-solid and fluid phases:
\begin{linenomath*}
\begin{eqnarray}
\begin{array}{lll}
\displaystyle{
\frac{\partial h}{\partial t} =
- \frac{\partial}{\partial x} \left [ h\lb \alpha_su_s + \alpha_{fs}u_{fs} + \alpha_{f}u_{f} \rb\right ]
- \frac{\partial}{\partial y} \left [ h\lb \alpha_sv_s + \alpha_{fs}v_{fs} + \alpha_{f}v_{f} \rb\right ]       },
\label{EnhancedGravity_h}
\end{array}    
\end{eqnarray}
\end{linenomath*}
where the hold up identity $\alpha_s + \alpha_{fs}+ \alpha_{f} = 1$ has been employed.
This way we can avoid the time derivatives in the terms associated with dispersion and enhanced gravity.

\section{Analysis of the Simplified Dispersion Relation}

Consider the dispersion for solid from (\ref{dispersion}):
{\footnotesize
\begin{linenomath*}
\begin{eqnarray}
\begin{array}{lll}
\displaystyle{\mathcal D_s^x = K_s^x\frac{h^2}{12}\left[ \lb U_s^2 
-\gamma^f_s \mathcal C^{s,f} \left( U_f^2-U_s^2\right)
-\gamma^{fs}_s \mathcal C^{s,fs} \lb U_{fs}^2-U_s^2\rb
\rb 
- \frac{D}{Dt} \lb U_s 
- \gamma^f_s\mathcal C^{s,f}\lb U_f - U_s \rb
- \gamma^{fs}_s\mathcal C^{s,fs}\lb U_{fs} - U_s \rb
\rb \right]} \\[3mm]
\displaystyle{
\hspace{.65cm}+K_s^x\frac{1}{\alpha_{s}}\frac{h^2}{6}\left [
C_{DG}^{s,f}|{\bf u}_f -{\bf u}_s|^{j-1} \lb U_f - U_s\rb
+C_{DG}^{s,{fs}}|{\bf u}_{fs} -{\bf u}_s|^{j-1} \lb U_{fs} - U_s\rb
-C_{DV}^s|{\bf u}_s| U_s\,\alpha_s
\right ].}
\label{Dispersion_A}
\end{array}    
\end{eqnarray}
\end{linenomath*}
}
\hspace{-3mm}
The flux in the momentum balance shows that in total the dispersion relation contains third order terms in flow depth, and third order derivatives of the flow velocities. These are the highest order terms therein. So, it is important to analyze the terms appearing in the dispersion relation, and additionally seek its simplifications and consequences. 

\subsection{The Role of Drag}

For the slowly varying slope parallel divergence, $U_s^2, U_{fs}^2, U_f^2$ can be neglected as compared to the other terms. Then, (\ref{Dispersion_A}) reduces to 
{\small
\begin{linenomath*}
\begin{eqnarray}
\begin{array}{lll}
\displaystyle{\mathcal D_s^x 
= -K_s^x\frac{h^2}{12}\left[
 \frac{D}{Dt} \lb U_s 
- \gamma^f_s\mathcal C^{s,f}\lb U_f - U_s \rb
- \gamma^{fs}_s\mathcal C^{s,fs}\lb U_{fs} - U_s \rb
\rb \right]} \\[3mm]
\displaystyle{
\hspace{.9cm}+K_s^x\frac{1}{\alpha_{s}}\frac{h^2}{6}\left [
C_{DG}^{s,f}|{\bf u}_f -{\bf u}_s|^{j-1} \lb U_f - U_s\rb
+C_{DG}^{s,{fs}}|{\bf u}_{fs} -{\bf u}_s|^{j-1} \lb U_{fs} - U_s\rb
-C_{DV}^s|{\bf u}_s| U_s\,\alpha_s
\right ].}
\label{Dispersion_B}
\end{array}    
\end{eqnarray}
\end{linenomath*}
}
For negligible virtual mass force, (\ref{Dispersion_B}) simplifies to
{\small
\begin{linenomath*}
\begin{eqnarray}
\begin{array}{lll}
\displaystyle{\mathcal D_s^x = -K_s^x\frac{h^2}{12}
 \frac{DU_s}{Dt}}
\displaystyle{+K_s^x\frac{1}{\alpha_{s}}\frac{h^2}{6}\left [
C_{DG}^{s,f}|{\bf u}_f -{\bf u}_s|^{j-1} \lb U_f - U_s\rb
+C_{DG}^{s,{fs}}|{\bf u}_{fs} -{\bf u}_s|^{j-1} \lb U_{fs} - U_s\rb
-C_{DV}^s|{\bf u}_s| U_s\,\alpha_s
\right ].}
\label{Dispersion_BC}
\end{array}    
\end{eqnarray}
\end{linenomath*}
}
Furthermore, for non-accelerating flows, the terms with $D/Dt$ vanish, and (\ref{Dispersion_BC}) further reduces to
\begin{linenomath*}
\begin{eqnarray}
\begin{array}{lll}
\displaystyle{\mathcal D_s^x = 
K_s^x\frac{1}{\alpha_{s}}\frac{h^2}{6}\left [
C_{DG}^{s,f}|{\bf u}_f -{\bf u}_s|^{j-1} \lb U_f - U_s\rb
+C_{DG}^{s,{fs}}|{\bf u}_{fs} -{\bf u}_s|^{j-1} \lb U_{fs} - U_s\rb
-C_{DV}^s|{\bf u}_s| U_s\,\alpha_s
\right ].}
\label{Dispersion_C}
\end{array}    
\end{eqnarray}
\end{linenomath*}
So, the interfacial and viscous drag may play an important role in generating dispersion relation in mixture mass flows which was not the case in the single-phase mass flows (Castro-Orgaz et al., 2015; Yuan et al., 2018).

\subsection{Negligible Dispersion}

In the most simple case the interfacial drags and the virtual mass may be neglected. A situation can arise such that the dispersion effect could be ignored. Then, from (\ref{Model_Final_Momentum_s}) and (\ref{Dispersion_A}), by integrating 
$\partial \left[\alpha_s h{\mathcal D_s^x}\right]/\partial x = 0$
with respect to $x$, we obtain:
\begin{linenomath*}
\begin{eqnarray}
\begin{array}{lll}
\displaystyle{ \frac{K_s^x}{12}\alpha_sh^3\lb U_s^2 - \frac{D U_s}{Dt}
-2 C_{DV}^s|{\bf u}_s| U_s
\rb} = {\mathcal P_{f_0}},
\label{Dispersion_D}
\end{array}    
\end{eqnarray}
\end{linenomath*}
where ${\mathcal P_{f_0}}$ is a constant of integration. However, determination of ${\mathcal P_{f_0}}$ may involve complex physical processes (explained in Section 4.3 - Section 4.5).
For simplicity, we assume a channelized flow, so the variation of the flow dynamic quantities with $y$ is negligible. For notational convenience we write $u = u_s$ and $\beta = C_{DV}^s$. Then, for $u_s > 0$, (\ref{Dispersion_D}) reduces to
\begin{linenomath*}
\begin{eqnarray}
\begin{array}{lll}
\displaystyle{\frac{\partial^2u}{\partial x\partial t} 
+\frac{\partial^2 \lb u^2/2\rb}{\partial x^2}
- 2\lb \frac{\partial u}{\partial x}\rb^2 
+\beta\frac{\partial}{\partial x}\lb u^2\rb
=  -{\mathcal P_{f}},}
\label{Dispersion_E}
\end{array}    
\end{eqnarray}
\end{linenomath*}
where 
\begin{linenomath*}
\begin{eqnarray}
\begin{array}{lll}
\displaystyle{{\mathcal P_f} = \frac{12{\mathcal P_{f_0}}}{K_s^x\alpha_sh^3}}.
\label{Lambda_1}
\end{array}    
\end{eqnarray}
\end{linenomath*}
We call ${\mathcal P_f}$ the (dissipative) prime-force coefficient (or, simply the ${\mathcal P}$-force coefficient).
Equation (\ref{Dispersion_E}) can be solved analytically only with some further assumptions. And, the solutions are presented in Section 4.3. If the solid particle distribution is uniform and the flow height can be approximated (by a constant), e.g., for a smooth flow, then, ${\mathcal P_{f}}$ is a constant.  
Equation (\ref{Dispersion_E}) can further be simplified as follows. 
\\[3mm]
{\bf I. Negligible $\lb \partial u/\partial x\rb^2$:} First, assume that $\partial u/\partial x$ is small and thus $\lb \partial u/\partial x\rb^2$ can be neglected. Then, integrating (\ref{Dispersion_E}) with respect to $x$, we obtain:
\begin{linenomath*}
\begin{eqnarray}
\begin{array}{lll}
\displaystyle{\frac{\partial u}{\partial t} 
+\frac{\partial \lb u^2/2\rb}{\partial x} 
= -\beta u^2-{\mathcal P_{f}} x + \alpha,}
\label{Dispersion_F}
\end{array}    
\end{eqnarray}
\end{linenomath*}
where $\alpha$ is a constant of integration, and we call $-{\mathcal P_{f}} x$ the prime-force (or, simply the ${\mathcal P}$-force), per unit mass. With this, we draw an important conclusion, that for spatially slowly varying velocity field, non-dispersive flows degenerate into an advective-dissipative system with a complex source term. Here, dissipation refers to the viscous dissipation due to the drag contribution $-\beta u^2$, and also $-{\mathcal P_{f}} x$, that we will elaborate later.
 When ${\mathcal P_{f_0}} \to 0$, or $h$ is large (enough) then ${\mathcal P_{f}} \to 0$. Alternatively, consider sufficiently small $x$. In both situations, ${\mathcal P_{f}} x$ is negligible, and (\ref{Dispersion_F}) becomes an inviscid, dissipative Burgers' equation developed by  Pudasaini and Krautblatter (2022):
\begin{linenomath*}
\begin{eqnarray}
\begin{array}{lll}
\displaystyle{\frac{\partial u}{\partial t} 
+\frac{\partial \lb u^2/2\rb}{\partial x} 
=  \alpha-\beta u^2}.
\label{Dispersion_F_1}
\end{array}    
\end{eqnarray}
\end{linenomath*}
From a simple physical consideration, following Pudasaini and Krautblatter (2022), $\alpha$ can represent the net driving force for the landslide motion, defined later at Section 5.1. So, (\ref{Dispersion_F}) can be viewed as the formal extension of the Pudasaini and Krautblatter (2022) landslide velocity equation, who also 
 constructed numerous exact analytical solutions for (\ref{Dispersion_F_1}), including simple to very sophisticated ones. 
 \\[3mm]
{\bf The Super Inviscid Dissipative Burgers' Equation:}
 There are two fascinating aspects of (\ref{Dispersion_F_1}). First, by setting the dispersion structure (which is internal to the new model developed here) to zero, we obtained the reduced equation of landslide motion without dispersion in Pudasaini and Krautblatter (2022). Second, the emergence of (\ref{Dispersion_F_1}) explicitly proves the consistency of our new model with dispersion.  
 However, when ${\mathcal P_{f}} x \neq 0$, (\ref{Dispersion_F}) is the extension of the inviscid, dissipative Burgers' equation in Pudasaini and Krautblatter (2022), for which, no exact analytical solutions have so far been developed. Yet, the model (\ref{Dispersion_E}) is more complex and general than (\ref{Dispersion_F}). For this reason, 
 we call (\ref{Dispersion_F}) the extension, and (\ref{Dispersion_E}) the super generalization of the inviscid, dissipative Burgers' equation. 
\\[3mm]
{\bf II. Time Independent Flows:} Second, assume a time-independent (steady state) flow. Then, from (\ref{Dispersion_E}) we have
\begin{linenomath*}
\begin{eqnarray}
\begin{array}{lll}
\displaystyle{\frac{\partial^2 }{\partial x^2}\lb u^2\rb
- 4\lb \frac{\partial u}{\partial x}\rb^2 
+2\beta\frac{\partial}{\partial x}\lb u^2\rb
= -2{\mathcal P_{f}}.}
\label{Dispersion_F_2}
\end{array}    
\end{eqnarray}
\end{linenomath*}
Since $\alpha_s, K_s^x$ and $h$ are positive, the nature of solution depends on the sign of ${\mathcal P_{f_0}}$ and its magnitude in ${\mathcal P_{f}}$ as given in (\ref{Lambda_1}). 

\subsection{Analytical Solutions}

\begin{figure}[t!]
\begin{center}
  \includegraphics[width=13cm]{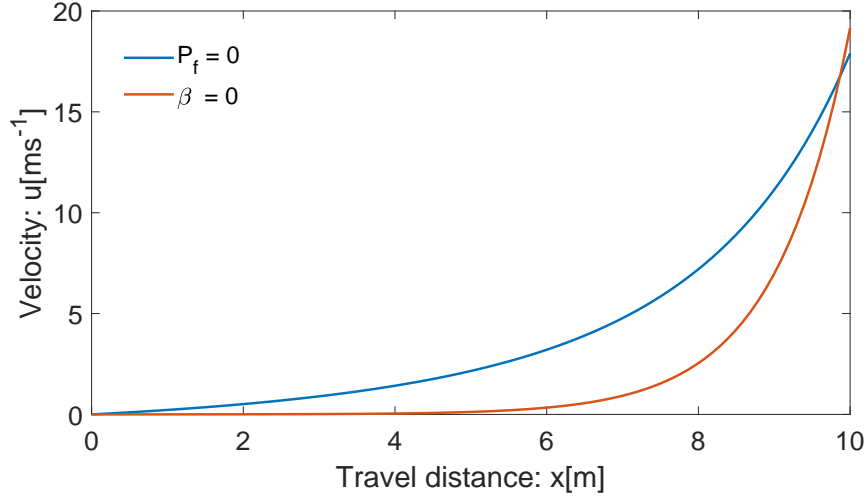}
  \end{center}
  \caption[]{Velocity fields represented by the solutions (\ref{Exact_1}) with parameters $C_1 = 0.05, C_2 = 1.0, \beta = 0.075$; and (\ref{Exact_2}) with parameters $C_1 = 0.01, C_2 = 0.005, {\mathcal P_{f}} = 2.5\times 10^{-6}$.}
  \label{Fig_1}
\end{figure}

Physically meaningful exact solutions explain the true and entire nature of the problem associated with the model equation (Pudasaini, 2011; Faug, 2015). The exact analytical solutions to simplified cases of non-linear debris avalanche model equations provide important insights into the full flow behavior of the complex system (Pudasaini and Krautblatter, 2022), and are often needed to calibrate and validate the numerical solutions (Pudasaini, 2016) as a prerequisite before running numerical simulations based on complex numerical schemes. So, such solutions should be developed, analyzed and properly understood prior to numerical simulations. This is very useful to interpret complicated simulations and/or avoid mistakes associated 
 with numerical simulations.
 Here, we construct some exact analytical solutions to (\ref{Dispersion_F_2}) for yet different simplified cases.
\\[3mm]
{\bf I. ${\mathcal P_{f}} = 0$, Vanishing Prime-force:} With this, the exact solution for (\ref{Dispersion_F_2}) takes the form:
\begin{linenomath*}
\begin{eqnarray}
\begin{array}{lll}
\displaystyle{
u(x) = C_2 \exp\left [{\frac{C_1 }{\beta}\exp(2 \beta x)}\right].
}
\label{Exact_1}
\end{array}    
\end{eqnarray}
\end{linenomath*}
There are two integration parameters $C_1, C_2$ to be determined, e.g., with the value and the slope of $u$ at a given point. 
\\[3mm]
{\bf II. $\beta = 0$, Vanishing Drag:} For this, the exact solution for (\ref{Dispersion_F_2}) becomes more complex:
\begin{linenomath*}
\begin{eqnarray}
\begin{array}{lll}
\displaystyle{
u(x) = \frac{\sqrt{-{\mathcal P_{f}}}\exp\lb -C_1\rb\tanh\left[ \exp\lb C_1\rb\lb C_2 +x\rb\right]}
{\sqrt{\tanh^2\left [ \exp\lb C_1\rb\lb C_2 +x\rb \right] -1}},
}
\label{Exact_2}
\end{array}    
\end{eqnarray}
\end{linenomath*}
where the two integration parameters $C_1, C_2$ are to be determined. The solutions (\ref{Exact_1}) and (\ref{Exact_2}) with some parameter values are presented in Fig. \ref{Fig_1} showing the exponential increase in the velocity field as a function of the travel distance. Where, for comparison the solution (\ref{Exact_1}) has been shifted down by about 2.
 However, more realistic solution is presented below when both ${\mathcal P_{f}}$ and $\beta$ cannot be ignored.
\\[3mm]
{\bf III. Small $\partial u/\partial x$:} Then, $\lb\partial u/\partial x\rb^2$ can be neglected in (\ref{Dispersion_F_2}) which, after integration, reduces to 
\begin{linenomath*}
\begin{eqnarray}
\begin{array}{lll}
\displaystyle{
u\frac{\partial u}{\partial x} = \alpha - \beta u^2 -{\mathcal P_{f}} x,}
\label{Dispersion_F_3}
\end{array}    
\end{eqnarray}
\end{linenomath*}
where $\alpha$ is a constant (the net driving force, see, Section 5.1), and 
\begin{linenomath*}
\begin{eqnarray}
\begin{array}{lll}
T^s_f = \alpha - \beta u^2 -{\mathcal P_{f}} x,
\label{Dispersion_F_3t}
\end{array}    
\end{eqnarray}
\end{linenomath*}
constitutes the total system force. The model (\ref{Dispersion_F_3}) includes both the parameters $\beta$ and ${\mathcal P_{f}}$ and extends the Pudasaini and Krautblatter (2022) landslide velocity equation for the time-independent motion for which their model corresponds to ${\mathcal P_{f}} = 0$. 
With the initial condition $u(0) = 0$, the exact analytical solution for (\ref{Dispersion_F_3}) yields:
\begin{linenomath*}
\begin{eqnarray}
\begin{array}{lll}
\displaystyle{
u(x) = \sqrt{\frac{\alpha}{\beta}}\sqrt{\left[ 1 - \exp\lb -2 \beta x\rb\right] - {\mathcal P_{u}} \left [ \lb 2 \beta x-1\rb + \exp\lb -2 \beta x\rb\right ]}
},
\label{Exact_3}
\end{array}    
\end{eqnarray}
\end{linenomath*}
where ${\displaystyle{{\mathcal P_{u}} = \frac{1}{2}\frac{1}{\alpha\beta}{\mathcal P_{f}}}}$. We call ${\mathcal P_{u}}$ the unified prime-force coefficient, which is a dimensionless number (quantity). It is induced by the prime-force coefficient ${\mathcal P_{f}}$, and also includes other force components, the net driving force $\alpha$, and the viscous resistance, represented by $\beta$.

\subsection{Postulation of the Prime-force: $-{\mathcal P_{f}} x$} 

The prime-force coefficient $-{\mathcal P_{f}}$ in (\ref{Dispersion_E}), and $x$ in the prime-force ${-\mathcal P_{f}} x$ in (\ref{Dispersion_F}) appear systematically. It emerged from our new modelling approach, with physical-mathematical foundation, from integrating the rate of acceleration, and the acceleration itself. This is exactly the reason why ${-\mathcal P_{f}} x$ is a dissipative (or anti-dissipative) force, and ${-\mathcal P_{f}}$ is the spatial rate of the prime-force along the slope. So, the new prime-force is physically meaningful. The values of ${\mathcal P_{f}}$ should be estimated with the dissipative processes taking place along the channel. It requires some extra and proper understanding of the flow dynamics to exactly determine ${\mathcal P_{f}}$ in (\ref{Dispersion_E}) and, thus, the force $-{\mathcal P_{f}} x$ itself. However, we have formally postulated (or invented) a new force mechanism, the prime-force $-{\mathcal P_{f}} x$, and have shown the physical ground for its existence. 
 Due to the presence of the term $-{\mathcal P_{f}} x$, the landslide velocity model (\ref{Dispersion_F_3}), and its solution (\ref{Exact_3}) are novel. 
The term $-{\mathcal P_{f}} x$ in (\ref{Dispersion_F_3}) adds some dissipative force that results in the deviation of the solution from the reference solution, ${\mathcal P_{f}} = 0$, produced by the driving force $\alpha$ and the viscous resistance associated with $\beta$. We can perceive $-{\mathcal P_{f}} x$ in different ways. It can be seen as the congregate of space dependent dissipative forces. Yet, $-{\mathcal P_{f}} x$ can be realized as any additional force other than the driving force $\alpha$ and the viscous resistance $-\beta u^2$ in their classical forms, which, unlike $-{\mathcal P_{f}} x$, do not contain any spatially varying dissipative contributions. As it is a completely new term and conception, its physical meaning and significance is worth exclusive elaboration in (\ref{Dispersion_E}), (\ref{Dispersion_F}), (\ref{Dispersion_F_2}), (\ref{Dispersion_F_3}), and (\ref{Exact_3}). 
As demonstrated below in Fig. \ref{Fig_2} and Fig. \ref{Fig_3}, the prime-force turned-out to be very useful in controlling the mass flow dynamics, or any other dynamical system, that can be described by the structure of the model equations presented here.

\subsubsection{Constraining ${\mathcal P_{f}}$} 

We need to physically constrain ${\mathcal P_{f}}$ in (\ref{Dispersion_E}). 
 Here, we present two possible scenarios. Without loss of generality, we impose physically legitimate and mathematically consistent conditions on the velocity and its derivatives at some position $x_0$ somewhere along the channel, or at appropriately chosen near source location. 
\\[3mm]
{\bf Scenario A:} First, consider plausible, but typical velocity and their gradients with magnitudes: $u\lb x_0\rb = 35, \lb\partial u/\partial x\rb \lb x_0\rb = 0.01, \lb\partial^2 u/\partial x^2\rb \lb x_0\rb = 0.00021$, and $\beta = 0.0019$. Then, from (\ref{Dispersion_E}), by neglecting the time variation of $\partial u/\partial x$, ${\mathcal P_{f}}$ assumes the value on the order of $-0.0085$ and ${\mathcal P_{u}} = -0.3$. However, similar values of ${\mathcal P_{f}}$ and ${\mathcal P_{u}}$ can be obtained with other physically admissible choices of $u\lb x_0\rb, \lb\partial u/\partial x\rb \lb x_0\rb, \lb\partial^2 u/\partial x^2\rb \lb x_0\rb$, and $\beta$.  
\\[3mm]
{\bf Scenario B:}
 Second, consider another plausible, but fundamentally different scenario, such that the velocity attains its local maximum somewhere at $x_0$ in the channel (e.g., a contracting flow). This is mathematically equivalent to $\lb\partial u/\partial x\rb \lb x_0\rb = 0$ and $\lb\partial^2 u/\partial x^2\rb \lb x_0\rb$ is negative, say $-0.00032$. With this, for the typical velocity of $u\lb x_0\rb = 35$, the estimated value of ${\mathcal P_{f}}$ is on the order of $0.0112$, and ${\mathcal P_{u}} = 0.4$. Again, similar values of ${\mathcal P_{f}}$ and ${\mathcal P_{u}}$ can be obtained with other physically admissible choices of $u\lb x_0\rb, \lb\partial u/\partial x\rb \lb x_0\rb$ and $\lb\partial^2 u/\partial x^2\rb \lb x_0\rb$.

\subsubsection{Dynamics of the Prime-force: $-{\mathcal P_{f}} x$}

\begin{figure}[t!]
\begin{center}
  \includegraphics[width=13cm]{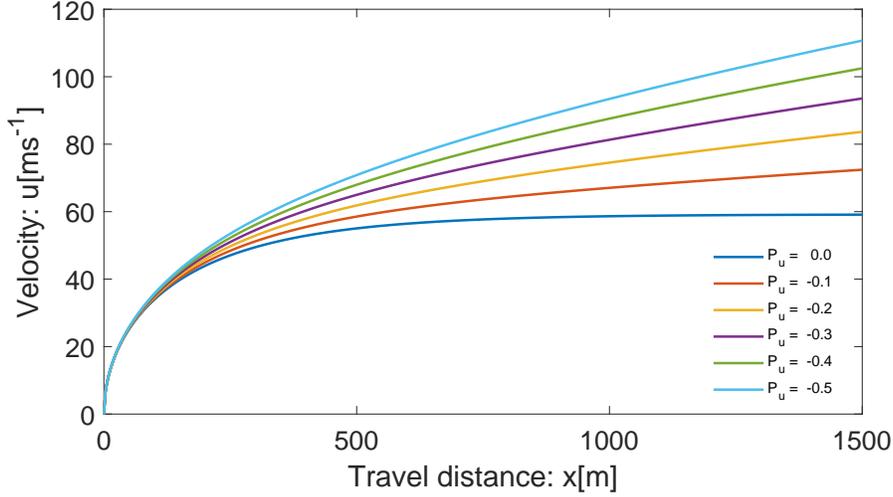}
  \end{center}
  \caption[]{The landslide motion enhanced by the prime-force $-{\mathcal P_{f}} x$, for ${\mathcal P_{f}} < 0$ given by the solution (\ref{Exact_3}), where ${\mathcal P_u} = {\mathcal P_{f}}/\lb 2 \alpha\beta\rb $.
  For any value of ${\mathcal P_{f}} < 0$, no matter how close it is to $0$, the system continuously deviates away from the reference state ${\mathcal P_{f}} = 0$, as long as ${\mathcal P_{f}} < 0$.}
  \label{Fig_2}
\end{figure}

 Solutions presented in Fig. \ref{Fig_2} for {\bf Scenario A}, with parameters $\alpha = 7.0$ and $\beta = 0.0019$ (as in Pudasaini and Krautblatter, 2022),
  show how the negative values of ${\mathcal P_{f}}$ (thus, the positive additional prime-force $-{\mathcal P_{f}} x$) enhances the motion from that discarding the effect of ${\mathcal P_{f}}$, i.e., ${\mathcal P_{f}} = 0$. As the value of ${\mathcal P_{u}}$ (or ${\mathcal P_{f}} = 2\alpha\beta\,{\mathcal P_{u}}$) decreases, the ${\mathcal P}$-force increases, and the velocity continuously deviates away from the reference (${\mathcal P_{f}} = 0$) state (solution). 
  Even a very small value of ${\mathcal P_{f}}$ pushes the system away from the reference state, and it continues to do so as ${\mathcal P_{f}}$ decreases. Thus, the term $-{\mathcal P_{f}} x$ with ${\mathcal P_{f}} < 0$ strongly weakens the drag force, adds to the pre-existing driving force, and thus the reference-state is never reached. It can be a possible scenario as the mass travels further downstream such that the drag force is always weaker than the net driving force and the additional force generated by the new term, $-{\mathcal P_{f}} x$, along the slope. This means that, as long as the condition $ \lb \alpha -{\mathcal P_{f}} x\rb > \beta u^2$ is satisfied, the system accelerates, always.
\begin{figure}[t!]
\begin{center}
  \includegraphics[width=13cm]{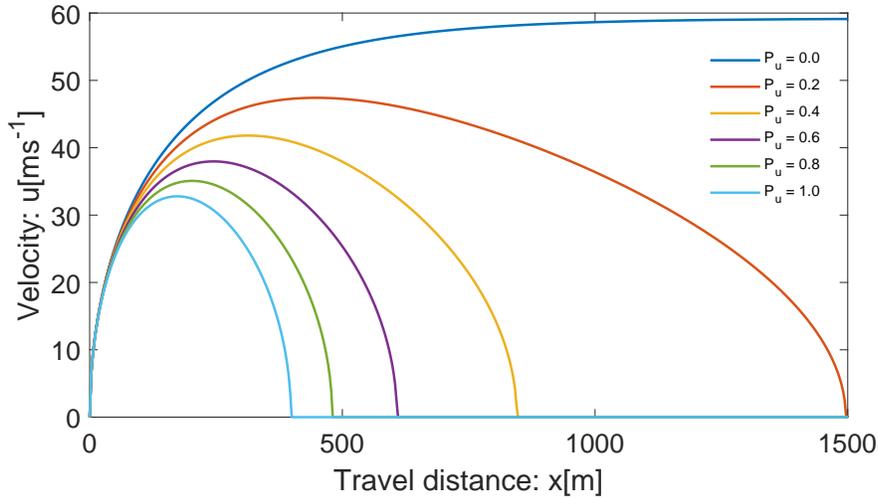}
  \end{center}
  \caption[]{The landslide motion controlled by the prime-force $ -{\mathcal P_{f}} x$, for ${\mathcal P_{f}} > 0$ given by the solution (\ref{Exact_3}), where 
  ${\mathcal P_u} = {\mathcal P_{f}}/\lb 2 \alpha\beta\rb $.
  The constrained velocity dome-curves and the reduced travel distances are shown.
  For any value of ${\mathcal P_{f}} > 0$, no matter how close it is to $0$, the system continuously bends below the reference state ${\mathcal P_{f}} = 0$, as long as ${\mathcal P_{f}} > 0$.}
  \label{Fig_3}
\end{figure}
\\[3mm]
Even more interesting, and perhaps physically more important, is the situation when ${\mathcal P_{f}} > 0$. 
This induces a spatially varying additional dissipative force resulting in the reduction of the total system force $T^s_f$ in (\ref{Dispersion_F_3t}) than before with the reference state, the solution with ${\mathcal P_{f}} = 0$, which effectively means that the mass decelerates as it slides downstream. This results in the reduced motion of the landslide. Then, depending on the magnitude of ${\mathcal P_{u}}$ (or ${\mathcal P_{f}} = 2\alpha\beta\,{\mathcal P_{u}}$), both the velocity and the travel distance will be reduced significantly to dramatically. The solutions are presented in Fig. \ref{Fig_3} for {\bf Scenario B}, with parameters $\alpha = 7.0$ and $\beta = 0.0019$, showing differently architectured beautiful dome-like constrained velocity fields and the firmly reduced mobility with increasing values of ${\mathcal P_{f}} > 0$. Interestingly, no matter how small, the novel observation is that, any positive value of ${\mathcal P_{f}}$ results in the significantly reduced mobility (velocity) and the run-out. This can happen, if there emerges any (other) energy dissipation mechanism along the slope. This effectively means that the total system force $T^s_f$ is continuously reduced as the mass slides downslope. So, after a certain position, the situation may prevail such that 
$ \beta u^2 > \lb \alpha -{\mathcal P_{f}} x\rb $, and the system decelerates along the slope, always, as long as ${\mathcal P_{f}} > 0$. This results in the reduced motion and the travel distance. 
\\[3mm]
 Both Fig. \ref{Fig_2} and Fig. \ref{Fig_3} demonstrate that the term $-{\mathcal P_{f}} x$ in (\ref{Dispersion_F_3}) can quickly and strongly compel the system away from its reference state (${\mathcal P_{f}} = 0$).
From the physical point of view, the ${\mathcal P}$-force $\lb-{\mathcal P_{f}} x\rb$ is associated with any possible spatially varying dissipative (or anti-dissipative) force. This may include any elements of forces that are not contained in $\alpha$ and $\beta$. The Coulomb-type force in $\alpha$ and the drag force associated with $\beta$ are almost exclusively used in mass flow simulations. However, the spatially dependent ${\mathcal P}$-force, postulated here, is entirely new, that was made possible with our modelling technique. Yet, as revealed by Fig. \ref{Fig_2} and Fig. \ref{Fig_3}, it helps to fundamentally and precisely control the dynamics, deposition and run-out of the landslide. We formally summarize these results in a Theorem.
\\[3mm]
{\bf The ${\mathcal P}$-force Theorem 4.1:} {\it There exists a unique number ${\mathcal P_{f}} > 0$ such that the landslide run-out (motion/dynamics) described by the dynamical equation}
\begin{linenomath*}
\begin{eqnarray}
\begin{array}{lll}
\displaystyle{\frac{\partial u}{\partial t} 
+\frac{\partial \lb u^2/2\rb}{\partial x} 
= \alpha -\beta u^2-{\mathcal P_{f}} x,}
\label{Dispersion_F_Th}
\end{array}    
\end{eqnarray}
\end{linenomath*}
{\it can be precisely controlled as expected. Here, $t$ is time, $x$ is the position along the slope, $u$ is the landslide velocity, $\alpha$ is the net driving force, $\beta$ is the viscous drag coefficient, and ${\mathcal P_{f}}$ is the prime-force coefficient.}

\subsection{The Prime-force: Essence, Implication and use in Simulation} 

Here, we further explain the essence and application potential of the new prime-force. Practitioners and applied researchers are frequently in trouble in controlling the motion and run-out of mass flows. One of the biggest problems in dealing with the natural mass flow events is the proper simulation of their flow velocities and the run-out distances. This also applies to industrial mass transports. We know that, more or less, until now, in real event simulations, the forces are used in a way that fits best to the data, sometimes very low (almost none) and sometimes substantially (much) higher than reality (Christen et al., 2010; Frank et al., 2015; Dietrich and Krautblatter, 2019; Mergili et al., 2020a, 2020b; Frimberger et al., 2021; Shugar et al., 2021). This clearly indicates that there are some physical processes operating in nature we were not aware of before. Now, we have formally proven that, in principle, such process exists, which can be quantified. The prime-force does exactly this by controlling the motion in a precise way. Our simple model, and particularly the emergence of the new prime-force, $-{\mathcal P_{f}} x$, can tremendously help to address this long standing problem. In this respect, the model (\ref{Dispersion_F_3}), and its exact analytical solution (\ref{Exact_3}), can be very useful for practitioners and engineers in efficiently and quickly simulating the motion of the landslide down the entire slope, accelerating and decelerating motions, and deposition as it comes to an standstill in a fully controlled manner.  
\\[3mm] 
There are two important aspects. ($i$) We have physically and mathematically proven that a new force structure, the prime-force, exists, which is extra to the known frictional or viscous forces. ($ii$) There are challenges related to the correct reproduction of field observations through simulations. Often, we have difficulties in adequately back-calculating the observed mass flow events. 
The prime-force is induced by the rate of spatially varying dissipative forces, but not merely the spatially varying friction and viscosity parameters.
So, the prime-force ($i$) will help to overcome the challenges in ($ii$) and accordingly support the practitioners. 
However, if it is only about the spatial distribution and evolution of friction and viscosity parameters, which we still do not at all understand, and also various numerical issues (e.g., cell size, topography and flow boundary), both do not involve the spatial rate of dissipative forces, the challenges in ($ii$) could still be addressed without the prime-force. 
\\[3mm]
 The Coulomb force cannot contain all the friction effects. The same applies to the viscous drag. As simulations often contrast the observations significantly, and none of the forces we know can reproduce the observation, there must be something extra to the Coulomb and viscous drag forces in the form we already know. The prime-force does exactly this. The prime force congregates all forces with spatially varying rate of dissipations that are not in $\alpha$ and $\beta$, and complement to what we know. The prime-force may even combine the Coulomb and viscous forces and generate a spatially varying rate of dissipation. 
One may yet think of producing similar results, as done above by the prime-force, by means of other forces which we know already. However, we can not achieve this by changing basal friction and/or the viscous drag. First, it is not possible in a classical way with Coulomb friction.
 The exact solution (\ref{Exact_3}) is constructed by assuming that $\alpha$ does not vary along the slope, while the ${\mathcal P}$-force, $-{\mathcal P_{f}} x$, by nature, does. The same is true for the drag force.
 Second, even by spatially varying the Coulomb friction (i.e., $\delta$) and/or the viscous drag ($\beta$), the motion, as controlled by the prime-force in (\ref{Dispersion_F_3}), cannot be achieved to precisely reproduce the observed run-out distance. Physically, $\delta$ is bounded from above, so often it is not able to control the motion in an appreciable way. Moreover, by definition, the viscous drag cannot bring the motion to a halt. But now, we can formally accommodate any additional energy dissipation mechanism in the ${\mathcal P}$-force accomplishing the observed effect rather than changing the Coulomb friction, whose value (as mentioned above) is often used arbitrarily in simulation to fit the data, or it does not exhibit any admirable effect. 
\\[3mm]
For granular, debris and particle-laden flows, several situations may arise where the dissipative (or anti-dissipative) force can increase (or decrease) as the mass moves downslope. There can be several factors aiding to the prime force. We mention some possible scenarios that may contribute to the spatial rate of the prime-force, i.e., ${\mathcal P_f}$. ($i$) Often the debris flow heads and lateral flanks become more and more granular dominated, or frictionally stronger due to phase-separation and/or particle sorting. These are observed phenomena (Johnson et al., 2012; de Haas et al., 2015, 2016; de Haas and van Woerkom, 2016; Pudasaini and Fischer, 2020b; Baselt et al., 2021, 2022). 
($ii$) The collisional and viscous dissipations can increase as flow moves on, e.g., by added particles and fines (the situation prevails due to basal erosion and entrainment), and increased agitations (de Haas et al., 2015, 2016; Pudasaini and Mergili, 2019; Pudasaini and Krautblatter, 2021). The viscous resistance can also increase due to added fragmented fine particles, e.g., in rock-ice avalanche motion (Pudasaini and Krautblatter, 2014). 
($iii$) The energy dissipation may increase in the downstream as the flow transits, e.g., from the glacial surface to the gravel-rich, or the rough moraine surface.
($iv$) Detailed topographic effects (Fischer et al., 2012), that could not be resolved otherwise, may also be included as an energy dissipation mechanism. 
\\[3mm]
In reality, the prime-force coefficient, ${\mathcal P_{f}}$, can be a complex function of some or all of those physical phenomena described above, and any other permissible circumstances associated with the dissipative mass flows with the rate of dissipative forces along the slope. 
Its admissible forms are yet to be determined. Still, ${\mathcal P_{f}}$ could also be constrained from laboratory experiments or from the field data with respect to the observed dynamics and the run-out. Alternatively, the practitioners may ascertain ${\mathcal P_{f}}$ in empirically adequate ways, if they prefer to do so.
This adds an additional uncertain parameter to the simulations, besides the existing ones. This may make parameter calibration and predictive simulations even more difficult, but helping to control the landslide as
observed.
However, we mention that, as the prime-force is a new concept, further intensive research would help to boost its clarity and expedite its practical applications.
\\[3mm]
Analytical solution presented in (\ref{Exact_3}) formally proves that the new dissipative force appreciably controls the motion and runout. Depending on its sign, it can enhance or control the motion, equivalently, stretch (Fig. \ref{Fig_2}) or reduce (Fig. \ref{Fig_3}) the travel distance (or coverage area). 
With this, we can now formally include the new dissipative force $-{\mathcal P_{f}} x$ (similarly in other directions) in the list of forces in the momentum balance equations (\ref{Model_Final_Momentum}), and implement the prime-force in any simulation of mass flow. There are some technical aspects to consider while implementing the new force in computing. 
$(i)$ Note that, ${\mathcal P_{f}}$ are relatively small numbers.
$(ii)$ In general, we can have different ${\mathcal P_{f}}$ for different phases. 
 $(iii)$ Because of the possible directional inhomogeneity, ${\mathcal P_{f}}$ can be different in $x$ and $y$ directions, say ${\mathcal P_{fx}}$ and ${\mathcal P_{fy}}$.
$(iv)$ We can formally include ${-\alpha_s\mathcal P_{fx}} x$ in the list of forces in (\ref{Source_x_s}), say at the end of it, similar for (\ref{Source_x_fs}) and (\ref{Source_x_f})  with ${\alpha_{fs}}$ and ${\alpha_{f}}$.
$(v)$ For the $y$-direction for solid, we should use ${-\alpha_s\mathcal P_{fy}} y$, but we should remember that the outward directions are the increasing directions. Similar for other phases in $y$-direction.
So, in principle, the prime-force can be relatively easily included in any computational softwares, such as the r.avaflow (Mergili and Pudasaini, 2021a,b)  in a straightforward way.

\section{A Simple Dispersion Equation}

Reducing the sophistication, we consider a geometrically two-dimensional motion down a slope. We further assume that the relative velocity between coarse and fine solid
particles $(u_s, u_{fs})$ and the fluid phase $(u_f)$ in the landslide (debris) material is negligible, that
is, $u_s \approx u_{fs} \approx u_f =: u$, and so is the viscous deformation of the fluid.
This means, for simplicity, we are considering an effectively single-phase mixture (consisting of solid particles composed of coarse solid and fine solid, and viscous fluid) flow (Pudasaini and Krautblatter, 2021, 2022).
Then, by summing up the mass and momentum balance equations in Section 2.2, we obtain a single mass and
momentum balance equation describing the motion of a landslide (or a mass flow) with the non-hydrostatic contribution as:
\begin{equation}
\frac{\partial h}{\partial t} +  \frac{\partial }{\partial x}\left ( hu\right ) = 0,
\label{Eqn_1}
\end{equation}
\begin{equation}
\frac{\partial}{\partial t}(hu) +  \frac{\partial }{\partial x} \left [ h \left \{ u^2 + \left (\alpha_s\beta_s +\alpha_s\beta_f \right)\frac{h}{2} +\left (\alpha_s\mathcal D_s + \alpha_f\mathcal D_f\right)\right\}\right ]= h S,
\label{Eqn_2}
\end{equation}
where,\\ \indent
$\alpha_f = \left (1-\alpha_s \right)$, 
$\displaystyle{
\alpha_s\beta_s +\alpha_f\beta_f = \left [ \left (1-\gamma_s^f \right)K_s\alpha_s + \left ( 1-\alpha_s\right)\right ]g^z
+ \left [ \alpha_s\left (K_s -1 \right)+1\right]
\lb \frac{D\bar w}{Dt} + C_{_{DV}}^s\bar w u \rb
}$,\\ \indent
 $\displaystyle{\alpha_s\mathcal D_s+\alpha_f\mathcal D_f = \frac{h^2}{12}\left [\alpha_s\left(K_s-1 \right) +1\right] \left\{ \left ( \frac{\partial u}{\partial x}\right )^2 
  -\frac{D}{Dt}\lb \frac{\partial u}{\partial x}\rb - 2C_{_{DV}}^s u\frac{\partial u}{\partial x}
  \right\}} $,\\  \indent
 $\displaystyle{S = g^x
  -\mu_s\alpha_s\left\{ \lb 1- \gamma_s^f\rb g^z + \frac{D\bar w}{Dt} + C_{_{DV}}^s\bar w u \right\}
  -\alpha_s \left\{\gamma_s^f g^z + \frac{D\bar w}{Dt} + C_{_{DV}}^s\bar w u\right\}\frac{\partial h}{\partial h}
  - C_{_{DV}}^su^2}$,\\[3mm]
  are the fluid fraction in the mixture, the coefficient emerging from the hydraulic pressure gradients for the solid and fluid including the enhanced effects due to non-hydrostatic contributions, the dispersion contributions emerging from the non-hydrostatic consideration, and the source containing the forces. 
  Together with the mass balance (\ref{Eqn_1}), the momentum balance (\ref{Eqn_2}) can be written as:
  {\small
 \begin{eqnarray}
&&\frac{\partial u}{\partial t} +  u\frac{\partial u}{\partial x}  
+ \left [ \left \{\left( \left( 1-\gamma_s^f\right)K_s + \gamma_s^f\right)\alpha_s+\left ( 1-\alpha_s \right )\right\}g^z
+\alpha_s \left\{ \lb \frac{\partial}{\partial t}+u \frac{\partial}{\partial x}\rb {\bar w} + C_{_{DV}}^s\bar w u \right\}
\right ] \frac{\partial h}{\partial x}
\nonumber\\
&&+ \frac{1}{h}\frac{\partial }{\partial x}\left[\left \{ \alpha_s\left (K_s -1 \right)+1\right\}\left [\frac{h^3}{12} \left\{ \left ( \frac{\partial u}{\partial x}\right )^2 -\frac{\partial}{\partial x}\frac{\partial u}{\partial t} - u \frac{\partial^2u}{\partial x^2}
-2C_{_{DV}}^s u \frac{\partial u}{\partial x}
\right\}
+\frac{h^2}{2} \left\{ \lb \frac{\partial}{\partial t}+u \frac{\partial}{\partial x}\rb {\bar w} + C_{_{DV}}^s\bar w u \right\}   \right]\right]\nonumber\\ 
&&= g^x
  -\mu_s\alpha_s\left[ \lb 1- \gamma_s^f\rb g^z + \left\{ \lb \frac{\partial}{\partial t}+u \frac{\partial}{\partial x}\rb {\bar w} + C_{_{DV}}^s\bar w u \right\} \right]
- C_{_{DV}}^su^2.
\label{Eqn_3}
\end{eqnarray}
}
\hspace{-3mm}
The second term on the left hand side of (\ref{Eqn_3}) describes the advection, while the third term (in the square bracket) describes the extent of the local deformation that stems from the hydraulic pressure gradient of the free-surface of the landslide in which $\left ( 1-\alpha_s\right)g^z\partial h/\partial x$ emerges from the hydraulic pressure gradient associated with possible interstitial fluids in the landslide, and the terms associated with ${\bar w}$ are the components from enhanced gravity. The fourth term on the left hand side are extra contributions in the flux due to the non-hydrostatic contributions. Moreover, the third term on the left hand side and the other terms on the right hand side in the momentum equation (\ref{Eqn_3}) represent all the involved forces. 
The first and second
terms on the right hand side of (\ref{Eqn_3}) are
 the gravity
acceleration, effective Coulomb friction that includes
lubrication $\left ( 1- \gamma_s^f\right )$, liquefaction $\left (
\alpha_s\right )$ (because, if there is no or substantially low amount of solid, the mass is fully
liquefied, e.g., lahar flows), the third term with $\bar w$ emerges from enhanced gravity, and the fourth term is the viscous drag,
respectively. Note that the term with $1-\gamma_s^f$ or $\gamma_s^f$ originates from the buoyancy effect. By setting $\gamma_s^f = 0$ and $\alpha_s = 1$, we obtain a dry landslide, grain flow, or an avalanche motion. However, we keep $\gamma_s^f$ and $\alpha_s$ also to include possible fluid effects in the landslide (mixture). 
\\[3mm]
Note that for $K_s = 1$ (which may prevail for extensional flows, Pudasaini and Hutter, 2007), the third term on the left hand side associated with $\partial h/\partial x$ simplifies drastically, because $\left \{ \left( \left( 1-\gamma_s^f\right)K_s + \gamma_s^f\right)\alpha_s+\left ( 1-\alpha_s \right )\right \}$ becomes unity.
 So, the isotropic assumption (i.e., $K_s = 1$) loses some important information about the solid content and the buoyancy effect in the mixture.

\subsection{A Landslide Dispersion Equation}

For simplicity, we introduce the notations as: $b = \left \{ \alpha_s\left (K_s -1 \right)+1\right\}$, $\alpha = \left [\,g^x-(1-\gamma_s^f)\alpha_s\mu_s g^z\,\right]$, and $\beta = C_{_{DV}}^s$. Here, $b, \alpha$ and $\beta$ are the pressure parameter, net driving force and the viscous drag coefficient, respectively. Assume that the time-dependent terms in (\ref{Eqn_3}) can be ignored in relation to other terms. Moreover, let $hu = {\mathcal F}$ be a typical flux, and $\partial u/\partial x$ is a small quantity such that $\lb \partial u/\partial x\rb^2$ is negligible. 
Consider the definition of $\bar w $ from (\ref{tau_zz_aaa}).
Then, with a long wave approximation (we suppose that $h$ can be approximated by a constant, or simply parameterize it, $h = h_0$), the momentum balance (\ref{Eqn_3}) can be reduced to yield a third-order inhomogeneous non-linear ordinary differential equation in $u$ with parameters ${\mathcal D_D}, {\mathcal D_{S1}}, {\mathcal D_{S2}}, \alpha, \beta$:
\begin{eqnarray}
{\mathcal D_P}\frac{\partial^3 u}{\partial x^3} + {\mathcal D_{S1}}\frac{\partial^2 u}{\partial x^2} + \lb u + {\mathcal D_{S2}}\rb \frac{\partial u}{\partial x}= \alpha - \beta u^2,
\label{Eqn_5}
\end{eqnarray}
where, $\displaystyle{{\mathcal D_P} = \pm\frac{1}{3} b h_0 {\mathcal F}}$, 
$\displaystyle{{\mathcal D_{S1}} = \pm\frac{1}{2}\left [ \frac{5}{6}\beta b h_0 + \mu_s\alpha_s\right]{\mathcal F}}$, 
${\displaystyle{{\mathcal D_{S2}} = \pm\frac{1}{2}\mu_s\beta\alpha_s{\mathcal F}}} $ are associated with dispersion. Here, the $\pm$ sign correspond to the primarily expanding or contracting flows, which can be obtained by separately analyzing the dispersive contributions in (\ref{Eqn_3}).
We call (\ref{Eqn_5}) the landslide dispersion equation in which ${\mathcal D_P}$ plays the primary role as it is associated with the highest order term therein, while ${\mathcal D_{S1}}$ and ${\mathcal D_{S2}}$ play the secondary role. So, ${\mathcal D_P}$ is termed as the prime dispersion parameter. This is a simple, yet very interesting, dispersion equation that characterizes the dispersion effect in the mass flow. 

\subsection{Solution to the Dispersion Equation (\ref{Eqn_5})}

\begin{figure}[t!]
\begin{center}
  \includegraphics[width=13cm]{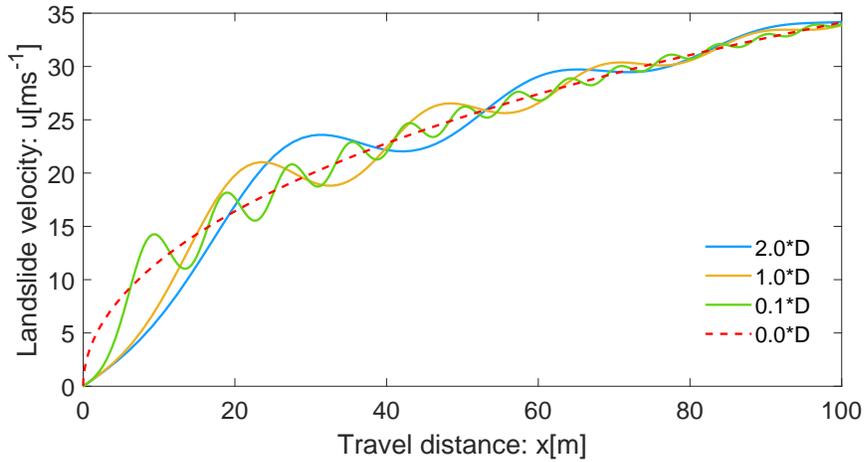}
  \end{center}
  \caption[]{The landslide velocity distribution with dispersion $\lb{\mathcal D_P = 327}, {\mathcal D_{S1} = 17}, {\mathcal D_{S2} = 0.03}\rb$ and without dispersion $\lb{\mathcal D_P} =0.0, {\mathcal D_{S1}} = 0.0, {\mathcal D_{S2}} = 0.0\rb$ described by (\ref{Eqn_5}). With dispersion ${\mathcal D}$, depending on its magnitude, the landslide behaves fundamentally differently by producing meanders of variable intensities around the reference state without dispersion.}
  \label{Fig_4}
\end{figure}

We analyze in detail the effect of dispersion in (\ref{Eqn_5}). Without the dispersive terms, (\ref{Eqn_5}) is the simple steady-state landslide velocity model developed in Pudasaini and Krautblatter (2022). We numerically solved (\ref{Eqn_5}) with the boundary conditions $\displaystyle{u(0) = 0.0, \frac{\partial u}{\partial x}(0) = 0.5, \frac{\partial^2 u}{\partial x^2}(0) = 0.0}$. The last two conditions are additionally required due to dispersion related dynamics. All conditions can be fixed based on the physics of the underlying problem. The results are shown in Fig. \ref{Fig_4} both with dispersion, ${\mathcal D_P = 327}, {\mathcal D_{S1} = 17}, {\mathcal D_{S2} = 0.03}$ (representing a realistic situation with $b = 1.0, h_0= 7.0, {\mathcal F} = 140, \alpha_s = 0.65, \mu_s = 0.36 \lb \delta_s = 20^\circ\rb, \alpha = 7.0, \beta = 0.0019$), and without dispersion ${\mathcal D_P = 0.0}, {\mathcal D_{S1} = 0.0}, {\mathcal D_{S2} = 0.0}$ effects. To demonstrate the influence of dispersion parameters ${\mathcal D}$ on the dynamics, we have amplified, downplayed, or ignored their values with different scales as $2.0\times{\mathcal D}, 1.0\times{\mathcal D}, 0.1\times{\mathcal D}, 0.0\times{\mathcal D}$, where the last value corresponds to the neglection of all dispersion effects. Figure \ref{Fig_4} clearly reveals fundamental effects of dispersion on the landslide dynamics. Moreover, the velocity distribution with dispersion is more complex due to its association with the higher-order derivative terms in (\ref{Eqn_5}). Dispersion produces a wavy velocity field of changing intensity about the simple reference state without dispersion. 
Local surge developments and attenuations as well as enhanced or hindered motions are often observed dynamical spectacles in landslides and debris avalanches. 
Such explicit description of the dispersive wave is the first of this kind for the avalanching debris mass. Once the landslide is triggered, the dispersive solution deviates significantly away from the non-dispersive one. However, after a sufficiently long distance, the dispersive solution tends to approach the non-dispersive state given by (\ref{Exact_3}) with ${\mathcal P_u = 0}$. Yet, significantly different scenarios can be generated with other sets of dispersion parameters. Alternatively, as ${\mathcal D_P \to 0.0}, {\mathcal D_{S1} \to 0.0}, {\mathcal D_{S2} \to 0.0}$, the dispersive wave coincides with the non-dispersive elementary solution. This proves the consistency of our model and also highlights the essence of dispersion in mass transport. 

 \subsection{Influence of the Solid Volume Fraction in Dispersion}
 
\begin{figure}[t!]
\begin{center}
  \includegraphics[width=13cm]{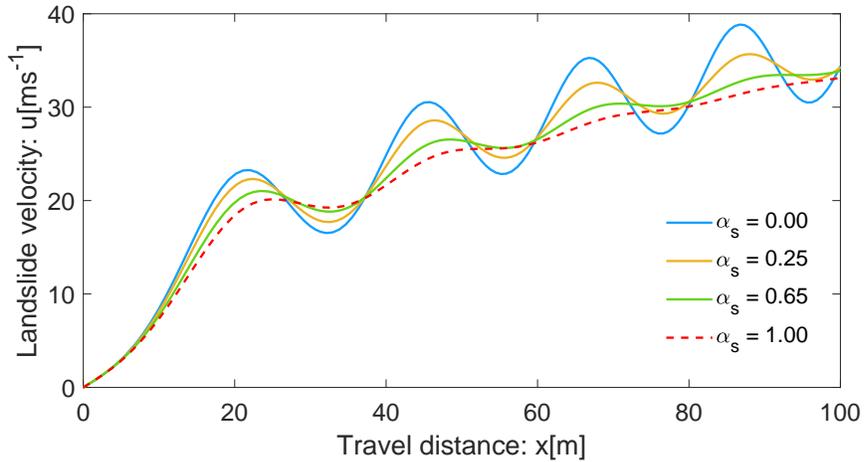}
  \end{center}
  \caption[]{Landslide velocity distributions with dispersion described by (\ref{Eqn_5}) for different solid volume fractions $\alpha_s$ in the landslide mixture. Dispersion increases firmly with decreasing solid volume fraction.}
  \label{Fig_5}
\end{figure}

 The solid volume fraction $\alpha_s$ is the key (physical) parameter in the mixture that governs the landslide motion and deformation. The strength of the landslide material is directly related to $\alpha_s$. The solid volume fraction influences all the parameters ${\mathcal D_P}, {\mathcal D_{S1}}, {\mathcal D_{S2}}$ and $\alpha$ in the dispersion equation (\ref{Eqn_5}), except $\beta$. So, here we analyze how the solid volume fraction regulates the landslide dispersion.
 Landslide velocity distributions with dispersion for different solid volume fractions in the mixture are presented in Fig. \ref{Fig_5}. 
 Dispersion is minimum for the fully dry material, and maximum for the vanishing solid fraction, akin to the fluid flow. The dispersion intensity increases energetically as the solid volume fraction decreases. This reveals that dispersion is related to the fluidness of the material. However, for higher values of $\alpha_s$ dispersion becomes weaker and weaker far downstream as compared to that near the source region.  
\begin{figure}[htb!]
\begin{center}
  \includegraphics[width=13cm]{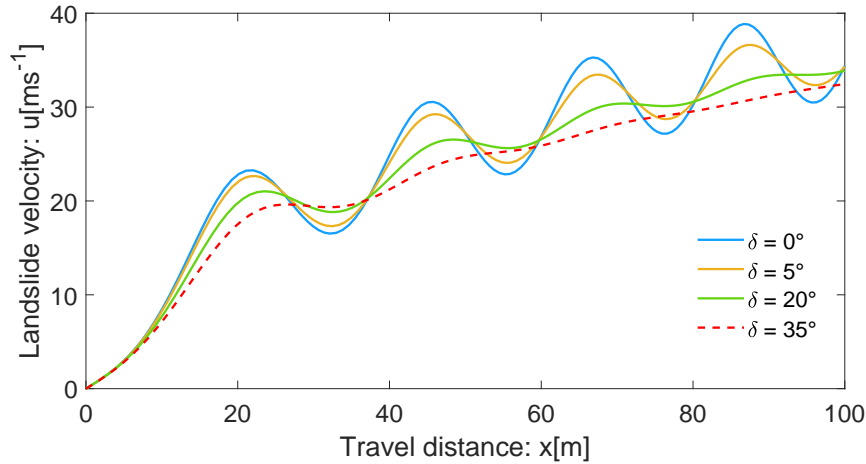}
  \end{center}
  \caption[]{Landslide velocity distributions with dispersion described by (\ref{Eqn_5}) for different basal friction angles $\delta$. Dispersion increases strongly with decreasing basal friction angle.}
  \label{Fig_6}
\end{figure}

\subsection{Influence of the Basal Friction in Dispersion}

The basal friction angle $\delta$ is a dominant physical parameter controlling the landslide dynamics. As for the solid volume fraction, the weaker material is associated with the lower friction angle. 
However, unlike the solid volume fraction, basal friction influences only ${\mathcal D_{S1}}, {\mathcal D_{S2}}$ and $\alpha$ in the dispersion equation (\ref{Eqn_5}), but not ${\mathcal D_P}$ and $\beta$. Landslide velocity distributions with dispersion for different frictions in the mixture are presented in Fig. \ref{Fig_6}. Dispersion increases strongly with decreasing values of $\delta$, with highest dispersion taking place for the motion of a frictionless material $\lb \delta = 0^\circ\rb$, akin to a fluid flow. However, for higher values of $\delta$, dispersion becomes relatively weaker as the landslide continues to propagate downstream.  
\\[3mm]
Both the solid volume fraction and the friction angle define the mechanical response of the landslide material against the applied forces, and govern the landslide motion and deformation. However, they regulate the landslide dynamics fundamentally differently, so are the dispersions with changing solid fractions and the basal frictions. These facts are demonstrated in Fig. \ref{Fig_5} and  Fig. \ref{Fig_6}. Although at the first glance, they look similar, the dispersion intensity is higher with the change of the basal friction as compared to that with the solid volume fraction. This can be explained, because basal fiction is the main physical parameter determining the landslide dynamics. These results are in line with our intuition and experience, and indicate the consistency of our model. This also sheds light on the physical significance of the simple dispersion model derived here.   

\section{Summary}

We considered the multi-phase mass flow model by Pudasaini and Mergili (2019) and extended it by including the non-hydrostatic contributions. This produces a novel non-hydrostatic multi-phase mass flow model. Effective normal stresses are constructed for all the solid, fine-solid and fluid phases in the mixture from the normal stress components, which include the interfacial momentum transfers such as the buoyancy, drag and virtual mass forces. Depending on the nature of the components in the effective normal stresses, the normal loads are separated into the enhanced gravity and the dispersion, which, respectively, correspond to the acceleration in the flow depth direction and the mass fluxes associated with the slope parallel directions. While drag and virtual mass forces appear in both, buoyancy is present only in the enhanced gravity for solid and fine-solid because it is associated with the reduced normal load of the solid particles in the mixture. As enhanced gravity and dispersion both emerge from the effective normal load, these enter into the lateral momentum fluxes via the hydraulic pressure gradients and additionally introducing the dispersion effects. This resulted in a complex and highly non-linear new contributions in the momentum fluxes. This may pose a great challenge in solving the model equations. This is mainly due to the involvements of the time derivatives in the fluxes that appear in the dispersion, and also in the enhanced gravity. 
To reduce the complexity, we have also presented some simplifications and approximations for the time derivatives appearing in the enhanced non-hydrostatic contributions. Similarly, we have presented analysis of the dispersion relations showing the role of the drag force. We discussed some special situations where the non-hydrostatic dispersive effects are more pronounced in multi-phase particle-fluid mixture mass flow than in single-phase flows. We proved that the negligible dispersion leads to the generalization of the existing inviscid, dissipative Burgers’ equation with source term.
We also presented simplified models that can help in solving the equations with reduced complexity. The reduced models already appeared to be the important generalizations and extensions of several mass flow models available in the literature. We formally postulated a novel, spatially varying dissipative (or anti-dissipative) force, called the prime-force. The practitioners and engineers may find the prime-force very useful in solving technical problems as it precisely controls the dynamics, run-out and the deposition of mass flows. We elucidated the need of formally including this new, physically-founded force in momentum balance equations.
We constructed a simple dispersion model and its solution that highlighted the essence of dispersion on the flow dynamics. We consistently demonstrated that dispersion produces a wavy velocity field around the reference state without dispersion. Our results show that dispersion increases strongly as the solid volume fraction and the basal friction decreases. The explicit description of dispersive waves and their control by the solid volume fraction and the basal friction are seminal understanding in mass flows. So, this contribution sets a foundation for a more complete and general simulation of non-hydrostatic dispersive, multi-phase mass flows.
\\[3mm]
 {\bf Acknowledgments:} Shiva P. Pudasaini acknowledges the financial support provided by the Technical University of Munich with the Visiting Professorship Program, and the international research project: AlpSenseRely $-$ Alpine remote sensing of climate‐induced natural hazards - from the Bayerisches Staatsministerium f\"ur Umwelt und Verbraucherschutz, Munich, Bayern.
 
\section*{Appendix}

\renewcommand{\theequation}{A.\arabic{equation}}

% reset the counter
\setcounter{equation}{0}

The expressions and discussions below are mainly based on Pudasaini and Mergili (2019).
\\[3mm]
{\bf A. The drag coefficients} are given by (Pudasaini, 2020): 
\begin{linenomath*}
\begin{subequations}\label{Drags}
\begin{align}
& \displaystyle{C_{DG}^{s,f} = \frac{\alpha_s \alpha_f\lb 1-\gamma_s^f\rb g}{\left [\mathcal U_T^{s,f}\left\{{\cal P}^{s,f}\mathcal F^{s,f}\lb Re_p^{s,f}\rb + \lb 1-{\cal P}^{s,f}\rb\mathcal G^{s,f}\lb Re_p^{s,f}\rb\right\}
+ {\mathcal S}_{\mathcal P}^{s,f}
\right ]^{\jmath}}},\label{Drags_s,f}\\[3mm] 
& \displaystyle{C_{DG}^{s,fs} = \frac{\alpha_s \alpha_{fs}\lb 1-\gamma_{s}^{fs}\rb g}{\left [\mathcal U_T^{s,fs}\left\{{\cal P}^{s,fs}\mathcal F^{s,fs}\lb Re_p^{s,fs}\rb + \lb 1-{\cal P}^{s,fs}\rb\mathcal G^{s,fs}\lb Re_p^{s,fs}\rb\right\}
+ {\mathcal S}_{\mathcal P}^{s,fs}
\right ]^{\jmath}}},\label{Drags_s,fs}\\[3mm] 
& \displaystyle{C_{DG}^{fs,f} = \frac{\alpha_{fs} \alpha_{f}\lb 1-\gamma_{fs}^f\rb g}{\left [\mathcal U_T^{fs,f}\left\{{\cal P}^{fs,f}\mathcal F^{fs,f}\lb Re_p^{fs,f}\rb + \lb 1-{\cal P}^{fs,f}\rb\mathcal G^{fs,f}\lb Re_p^{fs,f}\rb\right\}
+ {\mathcal S}_{\mathcal P}^{fs,f}
\right ]^{\jmath}}},\label{Drags_fs,f}
\end{align}
\end{subequations}
\end{linenomath*}
where, ${\cal P}^{s,f} \in (0, 1)$ is a function of the solid volume fraction ${\cal P}^{s,f} = \alpha_s^m$, {where $m$ is a positive number, close to 1},
combines the fluid-like, $\displaystyle{\mathcal F^{s,f} = {\gamma_s^f}\lb\alpha^f_s\rb^3 Re_p^{s,f}/180}$, and solid-like, $\displaystyle{\mathcal G^{s,f}= \alpha_f^{M^{s,f} -1}}$, drag contributions between solid and fluid components in three-phase mass flows; $\mathcal U_T^{s,f}$ is the terminal velocity of a particle falling through the fluid, $\jmath = 1 \, \mbox{or}\, 2$ is selected according to whether linear or quadratic drag coefficients are {used, and} $M^{s,f} = M^{s,f}\lb Re_p^{s,f}\rb$ depends on the particle Reynolds number $Re_p^{s,f} = \rho_fd_s~\mathcal U_T^{s,f}/\eta_f$ (Richardson and Zaki, 1954; Pitman and Le, 2005). Furthermore, $d_s$ is particle diameter, $\gamma_s^f = \rho_f/\rho_s$ is the fluid to solid density ratio, and $\alpha^f_s = \alpha_f/\alpha_s$ is the fluid to solid fraction ratio. 
 \\[3mm]
 $\displaystyle{{\mathcal S}_{\cal P}^{s,f} = \lb \frac{\mathcal P^{s,f}}{\alpha_s} + \frac{1 - {\mathcal P^{s,f}}}{\alpha_f}\rb {\mathcal K^{s,f}}}$ in (\ref{Drags_s,f}) is called the smoothing 
 function, where {${\mathcal K^{s,f}} = |\alpha_s{\mathbf u}_s + \alpha_f {\mathbf u}_f|$} is 
 determined by the mixture mass flux per unit mixture density, typically ${\mathcal K^{s,f}} = 10$ ms$^{-1}$ (Pudasaini, 2020). The emergence of $\mathcal S_{\cal P}^{s,f}$ in (\ref{Drags_s,f}) is crucial for the broad structure of the generalized drag that removes the singularity {from the existing drag coefficients (Pudasaini, 2020)}. With this, (\ref{Drags_s,f}) is called the enhanced generalized drag in mixture mass flows. This fully describes the drag for any values of the solid volume fraction $\alpha_s$. Similar discussions hold for the other drags ${C}_{DG}^{s,fs}$ and ${C}_{DG}^{fs,f}$. 
\\[3mm]
{\bf B. The virtual mass} induced mass and momentum enhancements for the solid-phase due to fluid and the fine-solid {are denoted by $u_s^{vm}$ and  $uu_s^{vm}$, $uv_s^{vm}$, and} are written as:
\begin{linenomath*}
\begin{subequations}\label{Virtual_mass_enhancements_s}
\begin{align}
& u_{s}^{vm} = \gamma_s^f {\mathcal C}^{s,f}\lb u_f -u_s \rb + \gamma_s^{fs} {\mathcal C}^{s,fs}\lb u_{fs} -u_{s} \rb ,\\[3mm] 
& uu_{s}^{vm}= \gamma_s^f {\mathcal C}^{s,f}\lb u_f^2 -u_s^2 \rb + \gamma_s^{fs} {\mathcal C}^{s,fs}\lb u_{fs}^2 -u_{s}^2 \rb,\\[3mm] 
& uv_{s}^{vm} = \gamma_s^f {\mathcal C}^{s,f}\lb u_{f}v_{f} -u_sv_s \rb + \gamma_s^{fs} {\mathcal C}^{s,fs}\lb u_{fs}v_{fs} -u_{s}v_{s}\rb.
\end{align}
\end{subequations}
\end{linenomath*}
The 
virtual mass force {coefficient} ${\mathcal C}^{s,f}$ in (\ref{Virtual_mass_enhancements_s}) is given by (Pudasaini, 2019): 
\begin{linenomath*}
\begin{eqnarray}
\displaystyle{ {\mathcal C}^{s,f} = \frac{\mathcal N_{vm}^0({\ell} + \alpha_s^n) - 1}{\alpha_s/\alpha_f + \gamma_s^f},
}
\label{VM_M}
\end{eqnarray}
\end{linenomath*}
where ${\mathcal N_{vm}}$ is the virtual mass number, and $\ell$ and $n$ are some numerical parameters. 
This model covers any distribution of the dispersive phase (dilute to dense distribution of the solid particles) that evolves automatically as a function of solid volume fraction. The physically most relevant values for the parameters 
can be: $\mathcal N_{vm}^0 = 10$, $\ell = 0.12$ and $n = 1$ (Pudasaini, 2019). The other virtual mass force coefficients ${\mathcal C^{s,fs}} $ and ${\mathcal C^{fs,f}}$ can be constructed from (\ref{VM_M}). Similarly, the virtual mass force induced mass and momentum enhancements for the fine-solid and fluid phases are given by:
 \begin{linenomath*}
\begin{subequations}\label{Virtual_mass_enhancements_fs1}
\begin{align}
& u_{fs}^{vm} = \gamma_{fs}^f  {\mathcal C}^{fs,f}\lb u_f - u_{fs}\rb - \alpha^s_{fs}{\mathcal C}^{s,fs}\lb u_{fs} - u_s\rb,\\[3mm] 
& uu_{fs}^{vm}= \gamma_{fs}^f  {\mathcal C}^{fs,f}\lb u_f^2 - u_{fs}^2\rb - \alpha^s_{fs}{\mathcal C}^{s,fs}\lb u_{fs}^2 - u_s^2\rb,\\[3mm] 
& uv_{fs}^{vm} = \gamma_{fs}^f  {\mathcal C}^{fs,f}\lb u_fv_f - u_{fs}v_{fs}\rb - \alpha^s_{fs}{\mathcal C}^{s,fs}\lb u_{fs}v_{fs} - u_sv_s\rb,
\end{align}
\end{subequations}
\end{linenomath*}
and
\begin{linenomath*}
\begin{subequations}\label{Virtual_mass_enhancements_f1}
\begin{align}
& u_{f}^{vm} = \alpha^s_{f} {\mathcal C}^{s,f}\lb u_f -u_s \rb + \alpha^{fs}_{f} {\mathcal C}^{fs,f}\lb u_{f} -u_{fs} \rb,\\[3mm] 
& uu_{f}^{vm}=  \alpha^s_{f} {\mathcal C}^{s,f}\lb u_f^2 -u_s^2 \rb + \alpha^{fs}_{f} {\mathcal C}^{fs,f}\lb u_{f}^2 -u_{fs}^2 \rb ,\\[3mm] 
& uv_{f}^{vm} = \alpha^s_{f} {\mathcal C}^{s,f}\lb u_fv_f -u_s v_s \rb + \alpha^{fs}_{f} {\mathcal C}^{fs,f}\lb u_{f} v_f -u_{fs} v_{fs}\rb,
\end{align}
\end{subequations}
\end{linenomath*}
 respectively, where, $\alpha^s_{fs} = \alpha_s/\alpha_{fs}, \alpha^s_{f} = \alpha_s/\alpha_{f}$ and $\alpha^{fs}_{f} = \alpha_{fs}/\alpha_{f}$ are the fraction ratios. 
 By consistently replacing $u$ by $v$ in (\ref{Virtual_mass_enhancements_s})-(\ref{Virtual_mass_enhancements_f1}), we obtain the virtual mass induced mass and momentum enhancements in the $y$-direction.
\\[3mm]
{\bf C. The $x$-directional fluid-type basal shear stresses in the $xz$-plane} are given, either by the no-slip condition (for both the fluid, and fine-solid):
\begin{linenomath*}
\begin{eqnarray}
\begin{array}{lll}\label{basal_shear_stress_u}
\displaystyle{
\left[\frac{\partial u_{f}}{\partial z}\right]_b = \chi_{u_{f}}\frac{u_{f}}{h}, \,\,\,
\left[\frac{\partial u_{fs}}{\partial z}\right]_b = \chi_{u_{fs}}\frac{u_{fs}}{h}
},
\end{array}
\end{eqnarray}
\end{linenomath*}
or by the no-slip condition for fluid, and the Coulomb-slip condition for fine-solid:
\begin{linenomath*}
\begin{eqnarray}
\begin{array}{lll}\label{basal_shear_stress_uu}
\displaystyle{
\left[\frac{\partial u_{f}}{\partial z}\right]_b = \chi_{u_{f}}\frac{u_{f}}{h}, \,\,\,
\left[\frac{\partial u_{fs}}{\partial z}\right]_b = \frac{C_{u_{fs}}^F}{\nu_{fs}^e}p_{fs} +2 C_{u_{fs}}^F \frac{\partial u_{fs}}{\partial x},}
\end{array}
\end{eqnarray}
\end{linenomath*}
with the Coulomb friction coefficient $C_{u_{fs}}^F = - u_{fs}/|{\bf u}_{fs}|\tan\delta_{fs}$, where $\delta_{fs}$ is the basal friction angle for the fine-solid.
The parameters $\chi_{u_f}$ and $\chi_{u_{fs}}$ in (\ref{basal_shear_stress_u}) and (\ref{basal_shear_stress_uu}) model the possible velocity distributions of the respective phases in the $xz$-plane normal to the sliding surface. 
\\[3mm]
{\bf D. The viscous stresses} associated with $\nu^e_{fs}$ and $\nu^e_{f}$ in (\ref{Source_x_fs})-(\ref{Source_x_f}) are related to the Newtonian-type viscous stresses. They include pressure, rate, yield strength and friction, see below.
\\[3mm]
{\bf E. The effective fluid and fine-solid kinematic viscosities} are given by:
\begin{linenomath*}
\begin{equation}\label{Viscosities}
\nu_f^e 
=  \nu_f + \frac{\tau_{y_f}}{||{\bf D}_f||}\left [ 1-\exp\lb-r_y||{\bf D}_f||\rb\right ],\,\,\, %\\[5mm]
\nu_{fs}^e 
=  \nu_{fs} + \frac{\tau_{y_{fs}}}{||{\bf D}_{fs}||}\left [ 1-\exp\lb-r_y||{\bf D}_{fs}||\rb\right ], 
\end{equation}
\end{linenomath*}
where $\tau_{y_f}$ and $\tau_{y_{fs}}$ are the {corresponding} yield stresses, $r_y$ are the parameters for regularization, and $\tau_{y_{fs}} = \sin\phi_{fs} p_{fs}$, 
and, {${\bf D}_f$ is the deviatoric strain-rate tensor for fluid}. 
In the viscosities (\ref{Viscosities}), {the} depth-averaged norm of ${\bf D}_f$ {is obtained as}:
\begin{linenomath*}
\begin{eqnarray}
\begin{array}{lll}
\displaystyle{{||\bf D}_f|| = {\left |4\frac{\partial u_f}{\partial x}\frac{\partial v_f}{\partial y} 
- \lb \frac{\partial u_f}{\partial y} + \frac{\partial v_f}{\partial x}\rb^2 
 - \lb \left [ \frac{\partial u_f}{\partial z}\right ]_b \rb^2
 - \lb \left [ \frac{\partial v_f}{\partial z}\right ]_b \rb^2
 \right |}^{1/2}},
 \label{Norm_D_f}
\end{array}
\end{eqnarray}
\end{linenomath*}
$||{\bf D}_f||$ is given by the second invariant ($I\!I_{{\bf D}_f}$) of the deviatoric strain-rate tensor for fluid: $||{\bf D}_f|| = \sqrt{I\!I_{{\bf D}_f}}$ with, $I\!I_{{\bf D}_f} = \displaystyle{\frac{1}{2}\left [ \text{tr}\lb {\bf D}_f\rb^2 - \text{tr}\lb {\bf D}_f^2\rb \right ]} $.
{The norm of the deviatoric strain-rate tensor for fine-solid, ${\bf D}_{fs}$, is obtained similarly.}
\\[3mm]
{\bf Flow and No-flow Regions:} 
The yield criteria help to precisely distinguish the flow and no-flow regions and depend on the rate of deformation and the material strengths for both the fine-solid and fluid phases (Prager and Drucker, 1952;  Domnik et al., 2013). Both the fine-solid and fluid phases yield plastically if the measures of the deviatoric stress tensors overcome the strengths of the materials. See, Pudasaini and Mergili (2019) for more details.
\\[3mm]
{\bf F. The $x$-directional enhanced non-Newtonian viscous stress} contribution (denoted by nN) for fine-solid due to the non-uniform distribution of the solid particles in the fine-solid is given by:
{\footnotesize
\begin{linenomath*}
\begin{align}\label{Non-Newtonian_fs}
\tau_{nN}^{{fs}^x} = \displaystyle{ \frac{\mathcal A^{fs,s}}{\alpha_{fs}}\left \{ 2\frac{\partial }{\partial x}\lb \nu_{fs}^e\frac{\partial \alpha_s}{\partial x}\lb u_{fs} - u_s\rb\rb 
 + \frac{\partial }{\partial y}\lb \nu_{fs}^e\lb\frac{\partial \alpha_s}{\partial x}\lb v_{fs} -v_s\rb + \frac{\partial \alpha_s}{\partial y}\lb u_{fs} - u_s\rb\rb\rb\right\}}
-\displaystyle{ \frac{\mathcal A^{fs,s}}{\alpha_{fs}}\frac{\xi_s\alpha_s\nu_{fs}^e\lb u_{fs} -u_s\rb}{h^2}}.
\end{align}
\end{linenomath*}
}
Similarly, the enhanced non-Newtonian viscous stress contribution for fluid due to the non-uniform distribution of the fine-solid and solid particles in the fluid is given by:
{\footnotesize
\begin{linenomath*}
\begin{align}\label{Non-Newtonian_f}
\tau_{nN}^{f^x} =  \displaystyle{ \frac{\mathcal A^{f,s}}{\alpha_f}\left \{ 2\frac{\partial }{\partial x}\lb \nu_f^e\frac{\partial \alpha_s}{\partial x}\lb u_f - u_s\rb\rb 
 + \frac{\partial }{\partial y}\lb \nu_f^e\lb\frac{\partial \alpha_s}{\partial x}\lb v_f - v_s\rb + \frac{\partial \alpha_s}{\partial y}\lb u_f - u_s\rb\rb\rb\right\}}
-\frac{\mathcal A^{f,s}}{\alpha_f}\frac{\xi_s\alpha_s\nu_f^e\lb u_f -u_s\rb}{h^2}\nonumber\\[1mm] 
 +\displaystyle{ \frac{\mathcal A^{f,fs}}{\alpha_f}\left \{ 2\frac{\partial }{\partial x}\lb \nu_f^e\frac{\partial \alpha_{fs}}{\partial x}\lb u_f - u_{fs}\rb\rb 
 + \frac{\partial }{\partial y}\lb \nu_f^e\lb\frac{\partial \alpha_{fs}}{\partial x}\lb v_f -v_{fs}\rb + \frac{\partial \alpha_{fs}}{\partial y}\lb u_f - u_{fs}\rb\rb\rb\right\}
-\frac{\mathcal A^{f,fs}}{\alpha_f}\frac{\xi_{fs}\alpha_{fs}\nu_f^e\lb u_f -u_{fs}\rb}{h^2}}.
\end{align}
\end{linenomath*}}

%{%\footnotesize
{\small

%}
 \end{document}